\documentclass[a4paper,12pt]{article}

\usepackage{amsthm}
\usepackage{amsmath}
\usepackage{amssymb}
\usepackage{latexsym}
\usepackage{authblk}
\usepackage{float}
\usepackage{diagbox} 
\usepackage{threeparttable}
\usepackage[textwidth=18cm,textheight=22cm]{geometry}
\usepackage[square,sort,comma,numbers]{natbib}

\newtheorem{theorem}{Theorem}[section]
\newtheorem{proposition}[theorem]{Proposition}
\newtheorem{lemma}[theorem]{Lemma}
\newtheorem{corollary}[theorem]{Corollary}
\newtheorem{remark}[theorem]{Remark}
\newtheorem{definition}[theorem]{Definition}
\newtheorem{example}[theorem]{Example}

\newcommand{\ud}{\mathrm{d}}
\newcommand{\eqdefr}{=\mathrel{\mathop:}}
\newcommand{\eqdefl}{\mathrel{\mathop:}=}
\newcommand{\Q}{\mathbb{Q}}
\newcommand{\FF}{\mathcal{F}}
\newcommand{\1}{\mathbf{1}}
\newcommand{\A}{\mathcal{A}}
\newcommand{\B}{\mathcal{B}}
\newcommand{\C}{\mathcal{C}}
\newcommand{\f}{\mathbb{F}}
\newcommand{\R}{\mathbb{R}}
\newcommand{\N}{\mathbb{N}}
\newcommand{\PP}{\mathbb{P}}
\newcommand{\Cov}{\mathbf{Cov}}
\newcommand{\Var}{\mathbf{Var}}
\newcommand{\Corr}{\mathbf{Corr}}

\newcommand{\ackname}{Acknowledgements}

\newcommand{\e}{\varepsilon}
\DeclareMathOperator{\inter}{int}
\DeclareMathOperator{\conv}{conv}
\DeclareMathOperator{\supp}{supp}
\DeclareMathOperator{\relint}{ri}
\DeclareMathOperator{\interior}{int}

\usepackage{graphicx}

\newcommand{\F}{\mathscr{F}}

\author{Hasanjan Sayit}

\affil{Xi'an Jiaotong Liverpool University, Suzhou, China}

\date{September 20, 2021}

\begin{document}
\title{Efficient frontiers for portfolios under SSD and law-invariant risk measures with hyperbolic return distributions}
\date{\today}

\maketitle

\abstract{In the classical Markowitz mean–variance framework, risk is measured by variance, and the portfolios on the efficient frontier can be derived in closed form using standard optimization methods. For broader mean–risk formulations, however, obtaining closed-form optimal portfolios is typically difficult. In this work, we derive explicit expressions for frontier portfolios corresponding to arbitrary law-invariant convex risk measures, assuming that the return vector follows a normal mean–variance mixture distribution. Our approach first establishes stochastic dominance relations within the family of normal mean–variance mixture models, and then leverages these relations to derive closed-form representations of frontier portfolios. The central finding demonstrates that when asset returns are described by a normal mean–variance mixture model, the associated mean–risk efficient frontier can be obtained by solving a Markowitz mean–variance problem for a suitably transformed return vector. The paper also extends the CAPM framework to the context of a normal mean–variance mixture distribution.}

\vspace{0.1in}


\vspace{0.1in}

\vspace{0.1in}

\section{Motivation}
In risk management and portfolio optimization, it is crucial to describe the underlying asset returns with appropriate probability distributions. The traditional Markowitz portfolio optimization framework relies on the assumption that asset returns follow a normal distribution. However, this normality assumption has been extensively challenged in the literature. A large body of empirical work on asset returns shows that models based on normality are often inconsistent with observed data. As the foundational study \cite{Eberlein_Ernst_And_Keller_Ulrich_1995} illustrates, empirical return densities exhibit higher concentration around the mean and in the tails, and lower mass in the intermediate regions, than the normal distribution. 

Another important property of empirical asset returns is their asymmetry, typically captured by skewness. Skewness in asset returns was identified early in the evolution of modern finance theory, and since then substantial work has focused on developing more refined models for return dynamics, motivated by the belief that improved models would enable better financial decisions. Consequently, it is clear that robust financial decision-making must be grounded in a well-specified multivariate probability distribution for asset returns.

Generalized hyperbolic distributions constitute a versatile family that can replicate most empirically observed characteristics of asset returns. As demonstrated in \cite{Eberlein_Ernst_And_Keller_Ulrich_1995}, hyperbolic distributions achieve an almost exact match to observed return density functions. Hyperbolic distributions are, in turn, a special case within the more general class of Normal mean-variance mixture (NMVM) models.

A $d$-dimensional random vector $X$ is said to follow a normal mean–variance mixture distribution with mixing variable $Z$ if, for each fixed $Z=z\in \mathbb{R}_+=(0,+\infty)$, the conditional distribution of $X$ is $X\mid_{Z=z}\sim W_d(\mu+z\gamma, z\Sigma)$. Here, $\mu,\gamma\in \mathbb{R}^d$ are fixed vectors, $\Sigma$ is a fixed positive-definite $d\times d$ matrix, and $W_d$ denotes a $d$-dimensional normal random vector with mean $\mu+z\gamma$ and covariance matrix $z\Sigma$. It is well known that such a random vector $X$ can be represented as
\begin{equation}\label{one}
X \overset{d}{=} \mu + \gamma Z + \sqrt{Z}\, A N_d,
\end{equation}
where $A=\Sigma^{1/2}$ and $N_d$ is a $d$-dimensional normal random vector with mean equal to the zero vector in $\mathbb{R}^d$ and covariance matrix given by the identity matrix $I_d$ in $\mathbb{R}^d\times \mathbb{R}^d$. The mixing variable $Z$ in (\ref{one}) is assumed independent of $N_d$. Throughout the paper we impose the condition 
\[
\mu+\gamma EZ\neq 0,
\]
on our model (\ref{one}).

Within the framework of (\ref{one}), several specifications have become particularly popular in financial modeling. When $Z$ is assumed to follow a Generalized Inverse Gaussian (GIG) distribution, the resulting distribution of $X$ is a multi-dimensional generalized hyperbolic (mGH) distribution. This family of distributions was originally introduced in \cite{Barndorff-1977intro}, and their extensive applications in finance have been explored in a number of subsequent works, including \cite{Aas_Kjersti_And_Haff_Ingrid_Hobaek_2006}, \cite{Eberlein_Ernst_2001}, \cite{Eberlein_Ernst_And_Keller_Ulrich_1995}, \cite{Madan_Dilip_B_And_Seneta_Eugene_1990}, \cite{Madan_Dilip_B_And_Carr_Peter_P_And_Chang_Eric_C_1998}, \cite{Eberlein-Raible}, and \cite{Eberlein-Kluge}. For instance, \cite{Madan_Dilip_B_And_Seneta_Eugene_1990} shows that the variance-gamma model and its elliptical multivariate extension provide an excellent description of asset returns over a unit time horizon. Similarly, \cite{Eberlein_Ernst_And_Keller_Ulrich_1995} demonstrates that hyperbolic distributions can deliver an almost perfect fit to empirical return density functions. In \cite{Aas_Kjersti_And_Haff_Ingrid_Hobaek_2006}, the authors find that the Generalized Hyperbolic Skew Student’s $t$ distribution matches empirical financial data extremely well.

In the multivariate context, \cite{Mcneil_A_J_And_Frey_R_And_Embrechts_P_2015} calibrates the multivariate generalized hyperbolic (mGH) model to both multivariate stock returns and multivariate exchange rate returns and shows, via likelihood-ratio tests, that the Gaussian specification is consistently rejected in favor of the more general mGH class. Furthermore, \cite{Aas_Kjersti_And_Haff_Ingrid_Hobaek_2006} employs multivariate NIG (Normal Inverse Gaussian) distributions in a risk management setting and demonstrates that they fit the empirical distribution of hedge fund returns substantially better than the Normal distribution. Generalized hyperbolic distributions have also been used in the valuation of interest rate derivatives, where they have been shown to yield highly accurate prices for caplets and other interest rate products; see, for example, \cite{Eberlein-Raible} and \cite{Eberlein-Kluge}.

In this paper, we examine a financial market consisting of $d$ risky assets and assume that their return vector $X$ satisfies (\ref{one}). The set of admissible portfolios is $\R^d$, and for any portfolio $\omega\in\R^d$ the associated portfolio return is $\omega^T X$, where $T$ denotes transposition. The central objective of this work is to analyze the optimization problem
\begin{equation}\label{asiddd}
\begin{split}
\min_{\omega}\; &\rho(-\omega^TX),\\
s.t.\; &E(-\omega^TX )=r,\\
&\omega^T\1=1,\\
\end{split}    
\end{equation}
for any prescribed target level of expected return $r\in\R$. In (\ref{asiddd}), $\rho$ represents a risk measure and $\1$ denotes the $d$-dimensional vector whose components are all equal to one. In this note, we derive explicit closed-form solutions to (\ref{asiddd}) for a class of risk measures $\rho$ satisfying certain assumptions that will be specified in Section 3.

One of these assumptions is that $\rho$ is law-invariant, meaning that $\rho(\eta_1)=\rho(\eta_2)$ whenever $\eta_1\overset{d}{=}\eta_2$. Using the relation
\[
\omega^TX\overset{d}{=}\omega^T\mu+\omega^T\gamma Z+\sqrt{Z}\omega^T\Sigma \omega N(0, 1),
\]
where $N(0, 1)$ is a standard normal random variable independent of $Z$, it follows that
\[
\rho(-\omega^TX)=\rho(-\omega^T\mu-\omega^T\gamma Z+\sqrt{Z}\omega^T\Sigma \omega N(0, 1)).
\]
We further impose that $\rho$ is compatible with second-order stochastic dominance. Consequently, as will become clear, the analysis of problem (\ref{asiddd}) can be reduced to studying stochastic dominance relations among one-dimensional NMVM models.

We consider an atomless probability space $(\Omega,\mathcal{F},P)$ on which all random variables appearing in this paper are defined. For $1\le k<\infty$, we write $L^k=L^k(\Omega,\mathcal{F},P)$ for the space of random variables whose $k$-th moment is finite. The space $L^{\infty}(\Omega,\mathcal{F},P)$ contains all essentially bounded random variables, and $L^{0}(\Omega,\mathcal{F},P)$ denotes the collection of all random variables taking finite values. Given vectors $\mathbf{x}=(x_1,\dots,x_d)^T$ and $\mathbf{y}=(y_1,\dots,y_d)^T$, their scalar product is defined by
$
\mathbf{x}\cdot\mathbf{y}=\mathbf{x}^T\mathbf{y}=\sum_{i=1}^d x_i y_i,
$
and the Euclidean norm of $\mathbf{x}$ is given by
$
|\mathbf{x}|=(x_1^2+\cdots+x_d^2)^{1/2}.
$
We denote by $\phi(\cdot)$ the probability density function of a standard normal random variable, and by $\varphi(\cdot)$ its cumulative distribution function.

Our original motivation for investigating problem (\ref{asiddd}) was to analyze its solutions under the widely used and popular risk measures $VaR$ and $CVaR$. The value-at-risk (VaR) is defined for any random variable in $L^{0}$. For $\eta \in L^0$, we denote by $q_{\alpha}^{-1}(\eta)=\inf\{a\in \R: F_{\eta}(a)\geq \alpha \}$ the left-continuous inverse of the cumulative distribution function $F_{\eta}(\cdot)$ of $\eta$. For each $\alpha \in (0, 1)$, the quantity $VaR_{\alpha}$ is then given by $VaR_{\alpha}(\eta)=q_{\alpha}^{-1}(\eta)$. The conditional value-at-risk (CVaR) is defined on $L^1$ by $CVaR_{\alpha}(\eta)=(1/(1-\alpha))\int_{\alpha}^1q_{\beta}(\eta)d\beta$ for any $\eta \in L^1$. 

In this note, we succeed in deriving a closed-form expression for (\ref{asiddd}) when $\rho = CVaR_{\alpha}$, whereas obtaining an explicit analytical solution for (\ref{asiddd}) in the case $\rho = VaR_{\alpha}$ seems to be challenging. The primary reason is that the risk measure $CVaR_{\alpha}$ is compatible with second-order stochastic dominance. Additional  properties of CVaR (although they are not used in this paper) include its continuity on $L^k$ for every $k \ge 1$; see Theorem 4.1 in \cite{Kaina-Ruschendrof} and page 11 of \cite{Shied}, together with the references cited therein, and its Lipschitz continuity with respect to $L^k$ convergence for $1 \le k \le \infty$; see Proposition 4.4 in \cite{WalterFarkas}. As already mentioned, the risk measure VaR is rigorously defined on $L^0$ and, in contrast to CVaR, is not a convex risk measure. It is compatible with first-order stochastic dominance but does not satisfy second-order stochastic dominance. These features make optimization problems involving VaR particularly challenging.

The rest of the paper is organized as follows. In Section 2, we investigate stochastic dominance relationships in one-dimensional NMVM models. Section 3 focuses on the analysis of frontier portfolios. In Section 4, we present the associated capital asset pricing model.

\section{High-order stochastic dominance}

Stochastic dominance (SD) relations between random variables are characterized through pointwise comparisons of their distribution functions. This notion plays a central role in economics and finance, with applications in option pricing, portfolio insurance, and risk management. Therefore, determining necessary and sufficient conditions for SD between random variables is highly relevant. In this section, we focus on deriving such conditions for higher-order SD relations within the class of one-dimensional NMVM models.

Let $U_k=\{U:\R \rightarrow \R : (-1)^{j-1}U^{(j)}(x)\ge 0,\ \forall x\in\R,\ 1\le j\le k\}$ denote the set of $k$-times differentiable utility functions whose derivatives alternate in sign, for each $k\ge 1$. We say that a scalar random variable $\eta_1$ \emph{stochastically dominates} another scalar random variable $\eta_2$ in the $n$th order if $EU(\eta_1)\ge EU(\eta_2)$ for all $U\in U_k$. Throughout the paper, we write $\eta_1 \succeq_{(k)} \eta_2$ to represent this SD relation.

This SD relation among random variables can also be characterized through their \emph{higher-order} cumulative distribution functions. For an arbitrary real-valued random variable $\eta$, let $F^{(1)}_{\eta}(x)$ denote its right-continuous distribution function. Then, for each integer $k \ge 2$, define recursively
\begin{equation*}
F^{(k)}_{\eta}(x)=\int_{-\infty}^x F^{(k-1)}_{\eta}(s)\,ds, \; \; \forall x\in \R.
\end{equation*}
We refer to $F^{(k)}_{\eta}(x)$ as the \emph{$k$th-order cumulative distribution function} of $\eta$, and we will use the shorthand notation $k$-CDF for brevity. For any two random variables $\eta_1$ and $\eta_2$, the relation $\eta_1\succeq_{(k)} \eta_2$ can be equivalently expressed as
\begin{equation}\label{sdn}
F^{(k)}_{\eta_1}(x)\le F^{(k)}_{\eta_2}(x), \; \; \forall x\in \R.
\end{equation}
If strict inequality in (\ref{sdn}) holds for at least one point $x$, we write $\eta_1\succ_{(k)} \eta_2$. Evidently, condition (\ref{sdn}) is intrinsically high dimensional, as the inequality must be verified for every real number $x$. Consequently, analyzing kSD relations among random variables is a nontrivial task. In this section, in order to obtain results on kSD relations for NMVM models, we verify condition (\ref{sdn}) directly, after deriving and studying the properties of their $k$-CDFs for all positive integers $k$. For additional discussion of these kSD dominance relations, we refer to \cite{Dentcheva-Ruszczynski}, \cite{Dent-Ruszczynski}, \cite{ogry-Ruszcz}, \cite{ogryczak-Ruszczyriski} and the references therein.

\subsection{High-order CDFs}

The main objective of this subsection is to derive explicit formulas for high-order CDFs corresponding to normal, elliptical, and NMVM models. Our computations in this part are largely based on key characterizations of the kSD property established in \cite{ogryczak-Ruszczyriski}. We begin by recalling several results from \cite{ogryczak-Ruszczyriski} that will be crucial for our analysis, and we adopt the same notation as in that paper.

For each $k \in \N$ (where $\N$ denotes the set of positive integers throughout) and any random variable $\eta \in L^k$, define
\[ 
G^{(k)}_{\eta}(x)=||(x-\eta)^+||_k, \quad \forall x\in \mathbb{R},
\]
where $x^+$ denotes the positive part of $x$. For any integer $k \geq 2$ and any pair $\eta_1, \eta_2 \in L^k$, Proposition 1 in \cite{ogryczak-Ruszczyriski} shows that kSD is equivalent to
\begin{equation*}\label{ksd}
G_{\eta_1}^{(k-1)}(x)\le G_{\eta_2}^{(k-1)}(x), \; \forall x\in \mathbb{R}. 
\end{equation*}

Furthermore, by Proposition 6 in \cite{ogryczak-Ruszczyriski}, the function $G_{\eta}^{(k)}(x)$ is increasing and convex, satisfies $\lim_{x\rightarrow -\infty}G_{\eta}^{(k)}(x)=0$, and obeys $G_{\eta}^{(k)}(x)\geq x-E\eta$ for all $x\in \R$. The same proposition also states that $x-E\eta$ is a right asymptote of $G_{\eta}^{(k)}(x)$, in the sense that
\[
\lim_{x\rightarrow +\infty}\left ( G_{\eta}^{(k)}(x)-(x-E\eta)\right ) =0.
\]

In the remainder of this section, we explicitly compute the $k$-CDFs for several classes of random variables for arbitrary $k \in \N$. These expressions are naturally linked to the kSD property via (\ref{sdn}).


\textbf{Normal random variables:} For a standard normal random variable $\eta \sim N(0, 1)$, we write $F_{\eta}^{(k)}(x)$ as $\phi^{(k)}(x)$. In particular,
\begin{equation*}\label{phik}
\phi^{(k)}(x)=\frac{1}{(k-1)!}E[(x-N)^+]^{k-1},    
\end{equation*}
where $N$ denotes a standard normal random variable. We now establish the following elementary lemma.
\begin{lemma}
For every integer $k \geq 2$, the following relation holds:
\begin{equation}\label{phinn}
\phi^{(k)}(x) = \frac{1}{k-1}\bigl[x\,\phi^{(k-1)}(x) + \phi^{(k-2)}(x)\bigr],
\end{equation}
where $\phi^{(0)}(x) = \phi(x)$ denotes the probability density function of a standard normal random variable, and $\phi^{(1)}(x) = \varphi(x)$ denotes its cumulative distribution function.
\end{lemma}
\begin{proof} We argue by induction. For $k=2$, we compute
\begin{equation*}
\begin{split}
\phi^{(2)}(x)
=&\int_{-\infty}^{x}\phi^{(1)}(s)\,ds
= s\phi^{(1)}(s)\big|_{-\infty}^{x}-\int_{-\infty}^x s\phi^{(0)}(s)\,ds\\
=&\,x\phi^{(1)}(x)+\phi^{(0)}(x),
\end{split}
\end{equation*}
where we used that $\lim_{s\rightarrow -\infty}[s\phi^{(1)}(s)]=\lim_{s\rightarrow -\infty}\phi^{(1)}(s)/\frac{1}{s}=0$, which follows from L'Hôpital's rule. Now suppose that (\ref{phinn}) holds for $k$; we show that it then holds for $k+1$. To this end, we integrate both sides of (\ref{phinn}) and afterwards substitute (\ref{phinn}). This yields
\begin{equation*}\label{26}
\begin{split}
\phi^{(k+1)}(x)
=&\int_{-\infty}^x\phi^{(k)}(s)\,ds
=\frac{1}{k-1}\int_{-\infty}^x s\phi^{(k-1)}(s)\,ds+\frac{1}{k-1}\phi^{(k-1)}(x)\\
=&\frac{1}{k-1}\big[s\phi^{(k)}(s)\big]\big|_{-\infty}^{x}-\frac{1}{k-1}\phi^{(k+1)}(x)+\frac{1}{k-1}\phi^{(k-1)}(x)\\
=&\frac{1}{k-1}\big[x\phi^{(k)}(x)\big]-\frac{1}{k-1}\phi^{(k+1)}(x)+\frac{1}{k-1}\phi^{(k-1)}(x),
\end{split}
\end{equation*}
where we have used $\lim_{s\rightarrow -\infty}[s\phi^{(k)}(s)]=\lim_{s\rightarrow -\infty}\phi^{(k)}(s)/\frac{1}{s}=0$, which follows from repeated applications of L'Hôpital's rule. Rearranging (\ref{26}) gives
\[
\phi^{(k+1)}(x)=\frac{1}{k}\big[x\phi^{(k)}(x)+\phi^{(k-1)}(x)\big],
\]
which establishes the induction step and completes the proof.
\end{proof}

\begin{remark}

From relation (\ref{phinn}) we obtain
\begin{equation*}\label{phinzero}
\phi^{(k)}(0)=\left \{
\begin{array}{ll}
\displaystyle \frac{1}{3\cdot 5\cdot 7\cdots (2i-3)\cdot (2i-1)} \,\frac{1}{\sqrt{2\pi}} & k=2i,\\[0.6em]
\displaystyle \frac{1}{4 \cdots (2i-2)\cdot (2i)}\,\frac{1}{2} & k=2i+1. \\
    \end{array}   
 \right.
\end{equation*}
We see that the sequence $\phi^{(k)}(0)$ is monotone decreasing and converges to zero. Moreover, from (\ref{phinn}) we deduce that $\phi^{(k)}(x)\geq \frac{x}{k-1}\phi^{(k-1)}(x)$, and hence, for $x\geq k-1$ it follows that $\phi^{(k)}(x)\geq \phi^{(k-1)}(x)$. Consequently, $\phi^{(k)}(x)$ and $\phi^{(k-1)}(x)$ must intersect at some point $x>0$.
\end{remark}

Next, using (\ref{phinn}), we obtain the following representation for $\phi^{(k)}(x)$.
\begin{lemma}
For any $k\geq 2$,
\begin{equation*}\label{phin}
\phi^{(k)}(x)=p_{k-1}(x)\varphi(x)+q_{k-2}(x)\phi(x),    
\end{equation*}
where $p_{k-1}(x)$ is a polynomial of degree $k-1$ satisfying
\begin{equation*}
p_j(x)=\frac{x}{j}p_{j-1}(x)+\frac{1}{j}p_{j-2}(x), \; \; j\geq 2,\;\; \; p_1(x)=x,\; \;  p_0(x)=1,    
\end{equation*}
and $q_{k-2}(x)$ is a polynomial of degree $k-2$ satisfying
\begin{equation*}
q_i(x)=\frac{x}{i+1}q_{i-1}(x)+\frac{1}{i+1}q_{i-2}(x),\; \;  i\geq 2, \;\; \; q_1(x)=\frac{x}{2},\; \;  q_0(x)=1.    
\end{equation*}
\end{lemma}

\begin{proof}
For $k=2$, (\ref{phinn}) immediately yields
\[
\phi^{(2)}(x)=x\varphi(x)+\phi(x)=p_1(x)\varphi(x)+q_0(x)\phi(x).
\]
For $k=3$, a direct computation gives
\[
\phi^{(3)}(x)=\frac{x^2+1}{2}\phi^{(2)}(x)+\frac{x}{2}\phi^{(0)}(x)
=p_2(x)\varphi(x)+q_1(x)\phi(x),
\]
again by applying (\ref{phinn}). Now we use induction. Assume it holds for all integers $2,3,\dots,k$, and we show it holds for $k+1$. From (\ref{phinn}), 
\begin{equation*}
\begin{split}
\phi^{(k+1)}(x)
&=\frac{x}{k}\phi^{(k)}(x)+\frac{1}{k}\phi^{(k-1)}(x)\\
&=\frac{x}{k}[p_{k-1}(x)\varphi(x)+q_{k-2}(x)\phi(x)]
+\frac{1}{k}[p_{k-2}(x)\varphi(x)+q_{k-3}(x)\phi(x)]\\
&=\bigg[\frac{x}{k}p_{k-1}(x)+\frac{1}{k}p_{k-2}(x)\bigg]\varphi(x)
+\bigg[\frac{x}{k}q_{k-2}(x)+\frac{1}{k}q_{k-3}(x)\bigg]\phi(x)\\
&=p_k(x)\varphi(x)+q_{k-1}(x)\phi(x),
\end{split}    
\end{equation*}
where we define
\[
p_k(x):=\frac{x}{k}p_{k-1}(x)+\frac{1}{k}p_{k-2}(x),\qquad
q_{k-1}(x):=\frac{x}{k}q_{k-2}(x)+\frac{1}{k}q_{k-3}(x).
\]
It is evident that $p_k(x)$ is a polynomial of degree $k$ and $q_{k-1}(x)$ is a polynomial of degree $k-1$.
\end{proof}

\begin{lemma} The function $y(x)=\phi^{(k)}(x)$ satisfies 
\begin{equation*}\label{y}
\begin{split}
&y''+xy'-(k-1)y=0,\\
&y(0)=\phi^{(k)}(0),\; y'(0)=\phi^{(k-1)}(0), \\  
\end{split}
\end{equation*}
and the polynomial solution $y(x)=\sum_{j=0}^{+\infty}a_jx^j$ of \eqref{y} is determined by
\[
a_{j+3}=\frac{(k-1)-(j+1)}{(j+2)(j+3)}a_{j+1},\; \;  j=0, 1, \cdots, 
\]
with initial coefficients
\[
a_2=\frac{k-1}{2}a_0,  \; \; a_1=\phi^{(k-1)}(0), \;\; a_0=\phi^{(k)}(0).
\]
\end{lemma}

\begin{proof} The relation \eqref{phinn} can be rewritten in the form
\[
(k-1)\phi^{(k)}(x)=x\phi^{(k-1)}(x)+\phi^{(k-2)}(x).
\]
Defining $y:=\phi^{(k)}(x)$, we have $y'=\phi^{(k-1)}(x)$ and $y''=\phi^{(k-2)}(x)$, so the differential equation follows immediately. To determine its polynomial solution, we substitute $y(x)=\sum_{j=0}^{+\infty}a_jx^j$, which yields a polynomial identically equal to zero. Consequently, each coefficient of this resulting polynomial must vanish, leading to the stated recurrence for $a_j$. 
\end{proof}

\begin{remark}\label{mre} Note that $a_{k+1}=0$, and thus $a_{k+2j+1}=0$ for every $j\geq 0$. Moreover,
\[
a_{k+2j}=(-1)^j\frac{3\cdot 5\cdots (2j-3)\cdot (2j-1)}{(k+2j)!}\frac{1}{\sqrt{2\pi}},
\]
and
\[
a_0=\phi^{(k)}(0),\; a_1=\phi^{(k-1)}(0),\; \cdots,\; a_j=\frac{1}{j!}\phi^{(k-j)}(0),\; \cdots,\; a_k=\frac{1}{k!}\phi^{(0)}(0)=\frac{1}{k!}\frac{1}{\sqrt{2\pi}}.
\]
\end{remark}\label{re26}

\begin{remark} In the case that $\eta \sim N(u, \sigma^2)$, we denote $F_{\eta}^{(k)}(x)$ by $\varphi^{(k)}(x; u, \sigma^2)$. We can easily obtain
\begin{equation*}\label{gphin}
\varphi^{(k)}(x; u, \sigma^2)=\sigma^{k-1}\phi^{(k)}(\frac{x-u}{\sigma}).    
\end{equation*}
By letting  $y(x)=\varphi^{(k)}(x; u, \sigma^2)$, one can easily show that it satisfies the following equation
\[
\sigma^2y^{''}+(x-u)y'-(k-1)y=0,
\]
with the initial conditions
\[
y(0)=\sigma^{k-1}\phi^{(k)}(-\frac{u}{\sigma}), \; y'(0)=\sigma^{k-2}\phi^{(k-1)}(-\frac{u}{\sigma}).
\]
\end{remark}

\textbf{Elliptical random variables:} Suppose $\eta \sim u+\sigma \sqrt{Z}N(0, 1)$, where $Z\in L^{k-1}$ is a positive random variable independent of $N(0, 1)$. In this case, we write $F^{(k)}_{\eta}(x)$ as $\varphi_e^{(k)}(x; u, \sigma)$. By (\ref{phinn}), we obtain
\begin{equation*}
 \varphi_e^{(k)}(x; u, \sigma)=\sigma^{k-1}\int_0^{+\infty}z^{\frac{k-1}{2}}\phi^{(k)}\!\Big(\frac{x-u}{\sigma \sqrt{z}}\Big)dF_Z(z).   
\end{equation*}

\begin{lemma}\label{27} For every $k\geq 2$, if $Z\in L^{k-1}$ then
\begin{equation*}\label{35}
\begin{split}
\frac{d\varphi_e^{(k)}(x; u, \sigma)}{du}=&-\sigma^{k-2}\int_0^{+\infty}z^{\frac{k-2}{2}}\phi^{(k-1)}\!\Big(\frac{x-u}{\sigma \sqrt{z}}\Big)dF_Z(z)<0,\\[0.2cm]
\frac{d\varphi_e^{(k)}(x; u, \sigma)}{d\sigma}=&\sigma^{k-2}\int_0^{+\infty}z^{\frac{k-1}{2}}\phi^{(k-2)}\!\Big(\frac{x-u}{\sigma \sqrt{z}}\Big)dF_Z(z)>0.
\end{split}
\end{equation*}
\end{lemma}
\begin{proof} The first identity  follows immediately. For the second one, observe that
\begin{equation*}\label{36}
\begin{split}
\frac{d\varphi_e^{(k)}(x; u, \sigma)}{d\sigma}=&(k-1)\sigma^{k-2}\int_0^{+\infty}z^{\frac{k-1}{2}}\phi^{(k)}\!\Big(\frac{x-u}{\sigma \sqrt{z}}\Big)dF_Z(z)\\
&+\sigma^{k-1}\int_0^{+\infty}z^{\frac{k-1}{2}}\Big(-\frac{x-u}{\sigma^2\sqrt{z}}\Big)\phi^{(k-1)}\!\Big(\frac{x-u}{\sigma \sqrt{z}}\Big)dF_Z(z)\\
=&\sigma^{k-2}\int_0^{+\infty}z^{\frac{k-1}{2}}\Big[(k-1)\phi^{(k)}\!\Big(\frac{x-u}{\sigma \sqrt{z}}\Big)-\Big(\frac{x-u}{\sigma\sqrt{z}}\Big)\phi^{(k-1)}\!\Big(\frac{x-u}{\sigma \sqrt{z}}\Big)\Big]dF_Z(z).
\end{split}
\end{equation*}
Using (\ref{phinn}), we obtain
\begin{equation*}\label{37}
(k-1)\phi^{(k)}\!\Big(\frac{x-u}{\sigma \sqrt{z}}\Big)-\frac{x-u}{\sigma \sqrt{z}}\phi^{(k-1)}\!\Big(\frac{x-u}{\sigma \sqrt{z}}\Big)=\phi^{(k-2)}\!\Big(\frac{x-u}{\sigma \sqrt{z}}\Big).
\end{equation*}
Making this substitution gives the second identity stated in the proposition.
\end{proof}

\textbf{NMVM random variables:}  When $\eta \sim a+b Z+\sigma \sqrt{Z}N(0, 1)$, we denote $F^{(k)}_{\eta}(x)$ by $\varphi_h^{(k)}(x; a, b, \sigma)$ (using the initial letter of “hyperbolic” distributions). From (\ref{phinn}), it follows that
\begin{equation*}\label{39}
\varphi_h^{(k)}(x; a, b, \sigma)=
\sigma^{k-1}\int^{+\infty}_0 z^{\frac{k-1}{2}}\phi^{(k)}\!\left(\frac{x-a-b z}{\sigma \sqrt{z}}\right)dF_Z(z).
\end{equation*}

By straightforward differentiation and by invoking (\ref{phinn}), we arrive at the following result.

\begin{lemma}\label{nc} For any integer $k\geq 2$, if $Z\in L^{k-1}$, then
\begin{equation}\label{40}
\begin{split}
\frac{d\varphi_h^{(k)}(x; a, b, \sigma)}{da}=&-\sigma^{k-2}\int_0^{+\infty}z^{\frac{k-2}{2}}\phi^{(k-1)}\!\left(\frac{x-a-bz}{\sigma \sqrt{z}}\right)dF_Z(z)<0, \\[4pt]    
\frac{d\varphi_h^{(k)}(x; a, b, \sigma)}{db}=&-\sigma^{k-2}\int_0^{+\infty}z^{\frac{k}{2}}\phi^{(k-1)}\!\left(\frac{x-a-b z}{\sigma \sqrt{z}}\right)dF_Z(z)<0,\\[4pt]
\frac{d\varphi_h^{(k)}(x; a, b, \sigma)}{d\sigma}=&\sigma^{k-2}\int_0^{+\infty}z^{\frac{k-1}{2}}\phi^{(k-2)}\!\left(\frac{x-a-b z}{\sigma \sqrt{z}}\right)dF_Z(z)>0.\\
\end{split}    
\end{equation}
\end{lemma}

\begin{proof} The first two identities in (\ref{40}) are obtained by differentiating 
the expression for $\varphi_h^{(k)}(x; a, b, \sigma)$ with respect to $a$ and $b$, respectively. The third identity follows by differentiating it with respect to $\sigma$ and then applying (\ref{phinn}) in the same way as in the proof of Proposition \ref{27}.
\end{proof}

\begin{remark} Lemma \ref{nc} implies that for any two random variables $\eta_1\sim a_1+b_1 Z+\sigma_1 \sqrt{Z}N(0, 1)$ and $\eta_2 \sim a_2+b_2 Z+\sigma_2 \sqrt{Z}N(0, 1)$, the condition
$a_1\geq a_2,\ b_1\geq b_2,\ \sigma_2\geq \sigma_1>0$ yields $\eta_1 \succeq_{(k)} \eta_2$. Moreover, if at least one of these inequalities is strict, then $\eta_1 \succ_{(k)} \eta_2$.
\end{remark}

\begin{remark} Denote the second-order CDF of $\eta$ in Lemma \ref{nc} by $F^{(2)}(a, b, \sigma)$. Then, by (\ref{39}), we obtain
\[
F^{(2)}(x; a, b, \sigma)=\int_0^{+\infty}\sigma \sqrt{z}\,\phi^{(2)}\!\left(\frac{x-a-b z}{\sigma \sqrt{z}}\right)dF_Z(z).
\]
For each fixed $x$, this is a differentiable function of the parameters $a, b, \sigma$. Thus, for fixed $x$ we can write
\[
dF^{(2)} = F^{(2)}_a\,da + F^{(2)}_b\,db + F^{(2)}_{\sigma}\,d\sigma,
\]
where $F^{(2)}_a, F^{(2)}_b, F^{(2)}_{\sigma}$ denote the partial derivatives. From Lemma \ref{nc} it follows that $F^{(2)}_a<0,\; F^{(2)}_b<0,\; F^{(2)}_{\sigma}>0$. Moreover, Lemma \ref{nc} provides the following explicit forms for these partial derivatives:
\begin{equation*}
 \begin{split}
 F_a^{(2)}=&-\int_0^{+\infty}\phi^{(1)}(\frac{x-a-bz}{\sigma \sqrt{z}})dF_Z(z),\\
  F_b^{(2)}=&-\int_0^{+\infty}z\phi^{(1)}(\frac{x-a-bz}{\sigma \sqrt{z}})dF_Z(z),\\
  F_{\sigma}^{(2)}=&\int_0^{+\infty}\sqrt{z}\phi^{(0)}(\frac{x-a-bz}{\sigma \sqrt{z}})dF_Z(z).\\ 
 \end{split}   
\end{equation*}
\end{remark}

\subsection{High-order SD}

As mentioned earlier, the condition (\ref{sdn}), which must be verified for all real numbers $x$, indicates that SD is an infinite-dimensional problem. Consequently, it is difficult to derive necessary and sufficient conditions for SD in general. For certain special classes of random variables, however, various characterizations of second-order stochastic dominance are well established in the literature. For instance, if $N_1\sim N(u_1,\sigma_1)$ and $N_2\sim N(u_2,\sigma_2)$, then $N_1\succeq_{(2)}N_2$ if and only if $u_1\ge u_2$ and $\sigma_1\le \sigma_2$; see, for example, Proposition 2.59 in \cite{Follmer-Schied(2004)} and Theorem 6.2 in \cite{HaimLevy}. For general random variables such convenient necessary and sufficient criteria for SD are, however, hard to obtain. In this section we provide some necessary and some sufficient conditions for kSD in the setting of NMVM models.

By definition, the $n_1$-th order SD relation is weaker than the $n_2$-th order SD relation whenever $n_1\ge n_2$, since $U_{n_1}\subset U_{n_2}$. Thus it is not obvious that, for normal random variables, $N_1\succeq_{(k)}N_2$ would still imply $u_1\ge u_2$ and $\sigma_1\le \sigma_2$ for $k\ge 3$. In this section, as a consequence of a more general result for NVMM, we prove that, in fact, $N_1\succeq_{(k+1)}N_2$ is equivalent to $u_1\ge u_2$ and $\sigma_1\le \sigma_2$ for every $k\in\N$; see Proposition \ref{2.77} below. Our argument provides an alternative proof of the classical second-order stochastic dominance characterization for normal random variables (cf. Theorem 6.2 in \cite{HaimLevy} and Theorem 3.1 in \cite{Dhompongsa2017StochasticDA}), and at the same time extends this type of characterization to kSD relations (in particular for $k\ge 3$) for elliptical random variables.

Before turning to the main calculations, we recall a pertinent result from \cite{ogryczak-Ruszczyriski}. With $G_{\eta}^{(k)}(x)$ defined as in Subsection 2.1, the value of $G_{\eta}^{(k)}(x)$ at $x=E\eta$ is called the central semideviation of $\eta$ and is denoted by $\bar{\delta}_{\eta}^{(k)}=G_{\eta}^{(k)}(E\eta)$ (we follow the notation of \cite{ogryczak-Ruszczyriski}). Explicitly,
\begin{equation*}\label{csd}
\bar{\delta}_{\eta}^{(k)} = \|(E\eta-\eta)^+\|_k = \bigl[E\bigl((E\eta-\eta)^+\bigr)^k\bigr]^{\frac{1}{k}}.
\end{equation*}
Proposition 4 in \cite{ogryczak-Ruszczyriski} states that $\bar{\delta}_{\eta}^{(k)}$ is convex, i.e.,
\[
\bar{\delta}_{t\eta_1+(1-t)\eta_2}^{(k)} \le t\bar{\delta}_{\eta_1}^{(k)} + (1-t)\bar{\delta}_{\eta_2}^{(k)}
\]
for every $t\in[0,1]$ and all $\eta_1,\eta_2\in L^k$. Furthermore, Corollary 2 of the same paper shows that if $\eta_1\succeq_{(k+1)}\eta_2$, then
\begin{equation}\label{key}
E\eta_1 - \bar{\delta}_{\eta_1}^{(n)} \ge E\eta_2 - \bar{\delta}_{\eta_2}^{(n)}, \;\;\; E\eta_1\ge E\eta_2,
\end{equation}
for all $n\ge k$, provided that $\eta_1,\eta_2\in L^n$. This relation (\ref{key}) will play a crucial role in the analysis carried out in this subsection. 

As a direct application of this result, we first establish the following proposition.
\begin{proposition}\label{2.77} Let $\eta_1 \sim u_1+\sigma_1 ZN(0, 1)$ and $\eta_2 \sim u_2+\sigma_2 ZN(0, 1)$ be two Elliptical random variables with $u_1, u_2\in \R$, $\sigma_1>0$, $\sigma_2>0$, and $Z$ a positive random variable satisfying $EZ^k<\infty$ for all positive integers $k$. Then, for every $k\geq 2$, we have $\eta_1 \succeq_{(k)} \eta_2$ if and only if $u_1\geq u_2$ and $\sigma_1\le \sigma_2$.
\end{proposition}
\begin{proof} The implication that $u_1\geq u_2$ and $\sigma_1\le \sigma_2$ yield $\eta_1 \succeq_{(k)} \eta_2$ for each $k\geq 2$ follows directly from Proposition \ref{27}. For the converse, observe that $E\eta_1=u_1$ and $E\eta_2=u_2$, and hence $u_1\geq u_2$ follows from Theorem 1 in \cite{ogryczak-Ruszczyriski}. To derive $\sigma_1\le \sigma_2$, note that the central semi-deviations of $\eta_1$ and $\eta_2$ are given by $\bar{\delta}^{(j)}_{\eta_1}=\sigma_1||(ZN)^+||_j$ and $\bar{\delta}^{(j)}_{\eta_2}=\sigma_2||(ZN)^+||_j$, respectively. Corollary 2 of \cite{ogryczak-Ruszczyriski} then implies
\begin{equation*}\label{38}
u_1-\sigma_1 ||(ZN)^+||_j\geq u_2-\sigma_2 ||(ZN)^+||_j,\quad \forall j\geq k.    
\end{equation*}
Since $(ZN)^+$ is an unbounded random variable, we have $\lim_{j\rightarrow \infty} ||(ZN)^+||_j=\infty$. Dividing both sides  by $||(ZN)^+||_j$ and letting $j\rightarrow \infty$, we obtain $-\sigma_1\geq -\sigma_2$, which is equivalent to $\sigma_1\le \sigma_2$. This completes the proof.
\end{proof}

Next, we examine kSD relations within the family of one–dimensional NMVM models. We begin by setting up some notation. Let $g(\cdot)$ be an arbitrary Borel measurable function from $(0,+\infty)$ to $(0,+\infty)$, and define
\[
Z_g := g(Z).
\]
For any real numbers $a_1,b_1,a_2,b_2$, set $\bar{X}_g = a_1 + b_1 Z_g + \sqrt{Z}\,N$ and $\bar{Y}_g = a_2 + b_2 Z_g + \sqrt{Z}\,N$. We first establish the following lemma.

\begin{lemma}\label{frac1}
Assume that the mixing variable $Z$ satisfies $Z \in L^k$ for every positive integer $k$. Then, for any bounded Borel function $g$ and any real numbers $a_1,b_1,a_2,b_2$, we have
\begin{equation*}\label{mlemma}
\lim_{k\to +\infty} \frac{\|(\bar{X}_g)^+\|_k}{\|(\bar{Y}_g)^+\|_k} = 1.
\end{equation*}
\end{lemma}

\begin{proof}
Note that
\[
\bar{X}_g - \bar{Y}_g = a_1 - a_2 + (b_1 - b_2) Z_g
\]
is a bounded random variable. Using $(a+b)^+ \le a^+ + b^+$ together with the triangle inequality for norms, we obtain
\begin{equation*}
\|(\bar{X}_g)^+\|_k = \|(\bar{X}_g - \bar{Y}_g + \bar{Y}_g)^+\|_k \le \|(\bar{X}_g - \bar{Y}_g)^+\|_k + \|(\bar{Y}_g)^+\|_k.
\end{equation*}
Hence,
\begin{equation*}\label{28}
\frac{\|(\bar{X}_g)^+\|_k}{\|(\bar{Y}_g)^+\|_k}
\le 1 + \frac{\|(\bar{X}_g - \bar{Y}_g)^+\|_k}{\|(\bar{Y}_g)^+\|_k}.
\end{equation*}
Since $(\bar{X}_g - \bar{Y}_g)^+$ is bounded, we have $\sup_{k\ge 1}\|(\bar{X}_g - \bar{Y}_g)^+\|_k < \infty$, while $(\bar{Y}_g)^+$ is unbounded, so $\|(\bar{Y}_g)^+\|_k \to \infty$ as $k\to\infty$. It follows that
\begin{equation*}\label{m1lemma}
\overline{\lim}_{k\to\infty}\frac{\|(\bar{X}_g)^+\|_k}{\|(\bar{Y}_g)^+\|_k} \le 1.
\end{equation*}

Applying the same reasoning in the reverse direction, we get
\begin{equation*}
\|(\bar{Y}_g)^+\|_k = \|(\bar{Y}_g - \bar{X}_g + \bar{X}_g)^+\|_k \le \|(\bar{X}_g - \bar{Y}_g)^+\|_k + \|(\bar{X}_g)^+\|_k.
\end{equation*}
Thus,
\begin{equation*}\label{32}
\frac{\|(\bar{X}_g)^+\|_k}{\|(\bar{Y}_g)^+\|_k}
\ge \frac{\|(\bar{X}_g)^+\|_k}{\|(\bar{X}_g - \bar{Y}_g)^+\|_k + \|(\bar{X}_g)^+\|_k}
= \frac{1}{\|(\bar{X}_g - \bar{Y}_g)^+\|_k/\|(\bar{X}_g)^+\|_k + 1}.
\end{equation*}
Since $(\bar{X}_g)^+$ is unbounded, we have $\|(\bar{X}_g)^+\|_k \to \infty$ as $k\to\infty$, and therefore
\[
\frac{\|(\bar{X}_g - \bar{Y}_g)^+\|_k}{\|(\bar{X}_g)^+\|_k} \to 0
\quad\text{as } k\to\infty.
\]
Consequently,  we obtain
\begin{equation*}\label{m2lemma}
\underline{\lim}_{k\to\infty}\frac{\|(\bar{X}_g)^+\|_k}{\|(\bar{Y}_g)^+\|_k} \ge 1.
\end{equation*}
Combining all these yields the result.
\end{proof}
For any bounded, positive Borel function $g$ and real numbers $a_1,a_2,b_1,b_2$ with $c_1>0$ and $c_2>0$, define
\[
\tilde{X}_g = a_1 + b_1 Z_g + c_1 \sqrt{Z}\,N,\qquad
\tilde{Y}_g = a_2 + b_2 Z_g + c_2 \sqrt{Z}\,N.
\]
The next proposition is the main result of this section.

\begin{proposition}\label{lembar}
Consider model (\ref{one}) and suppose that $Z \in L^k$ for every positive integer $k$. Then, for any positive bounded Borel function $g$ and each $k\in\mathbb{N}$, the relation $\tilde{X}_g \succeq_{(k+1)} \tilde{Y}_g$ implies
\[
a_1 + b_1 E Z_g \;\ge\; a_2 + b_2 E Z_g
\quad\text{and}\quad
c_1 \le c_2.
\]
\end{proposition}

\begin{proof}
Since
\[
E\tilde{X}_g = a_1 + b_1 E Z_g,\qquad
E\tilde{Y}_g = a_2 + b_2 E Z_g,
\]
the inequality $a_1 + b_1 E Z_g \ge a_2 + b_2 E Z_g$ follows directly from Theorem 1 in \cite{ogryczak-Ruszczyriski}. To establish $c_1 \le c_2$, we invoke Corollary 2 of the same reference \cite{ogryczak-Ruszczyriski}.

First, note that for each integer $j>k$ the central semi-deviation of order $j$ is given by
\[
\bar{\delta}_{\tilde{X}_g}^{(j)}
= c_1 \left\| \Big(\frac{b_1}{c_1} E Z_g - \frac{b_1}{c_1} Z_g + ZN\Big)^+ \right\|_j,
\qquad
\bar{\delta}_{\tilde{Y}_g}^{(j)}
= c_2 \left\| \Big(\frac{b_2}{c_2} E Z_g - \frac{b_2}{c_2} Z_g + ZN\Big)^+ \right\|_j.
\]
Set
\[
D_j := \left\| \Big(\frac{b_1}{c_1} E Z_g - \frac{b_1}{c_1} Z_g + ZN\Big)^+ \right\|_j,
\qquad
E_j := \left\| \Big(\frac{b_2}{c_2} E Z_g - \frac{b_2}{c_2} Z_g + ZN\Big)^+ \right\|_j.
\]
By Lemma \ref{frac1} we have $\lim_{j\to\infty} D_j/E_j = 1$. Moreover, since the random variables
\[
\Big(\frac{b_1}{c_1} E Z_g - \frac{b_1}{c_1} Z_g + ZN\Big)^+,
\quad
\Big(\frac{b_2}{c_2} E Z_g - \frac{b_2}{c_2} Z_g + ZN\Big)^+
\]
are unbounded, it follows that $\lim_{j\to\infty} D_j = +\infty$ and $\lim_{j\to\infty} E_j = +\infty$.

Corollary 2 in \cite{ogryczak-Ruszczyriski} then yields
\begin{equation*}\label{47}
a_1 + b_1 E Z_g - c_1 D_j \;\ge\; a_2 + b_2 E Z_g - c_2 E_j,
\end{equation*}
for all $j \ge k$. Dividing both sides by $E_j$ and letting $j\to\infty$, we obtain $-c_1 \ge -c_2$, which is equivalent to $c_1 \le c_2$.
\end{proof}

Now, for any positive Borel function $g$, any real numbers $b_1,b_2$, and any $c_1>0, c_2>0$, define
\[
\hat{X}_g = b_1 Z_g + c_1 \sqrt{Z}N,\qquad
\hat{Y}_g = b_2 Z_g + c_2 \sqrt{Z}N.
\]
We obtain the following corollary.

\begin{corollary}\label{iffandiff}
Assume that $Z \in L^k$ for all positive integers $k$. Then, for any positive, bounded Borel function $g$ and each $k\in\mathbb{N}$,
\[
\hat{X}_g \succeq_{(k+1)} \hat{Y}_g
\quad\Longleftrightarrow\quad
b_1 \ge b_2,\;\; c_1 \le c_2.
\]
\end{corollary}

\begin{proof}
If $\hat{X}_g \succeq_{(k+1)} \hat{Y}_g$, then Proposition \ref{lembar} gives $b_1 E Z_g \ge b_2 E Z_g$ and $c_1 \le c_2$. Since $E Z_g > 0$, this is equivalent to $b_1 \ge b_2$ together with $c_1 \le c_2$.

For the converse, consider
\[
Y_g := b Z_g + \sigma \sqrt{Z}N,\qquad b\in\mathbb{R},\; \sigma>0.
\]
Then
\[
\varphi_{Y_g}^{(k)}(x;b,\sigma)
=
\sigma^{k-1}\int_0^{+\infty} z^{\frac{k-1}{2}}
\phi^{(k)}\!\left(\frac{x - b z_g}{\sigma\sqrt{z}}\right) dF_Z(z),
\]
where $z_g = g(z)$. Proceeding as in the proof of Proposition \ref{nc}, we obtain
\[
\begin{split}
\frac{d}{db}\varphi_{Y_g}^{(k)}(x;b,\sigma)
&=
-\,\sigma^{k-2}\int_0^{+\infty} z^{\frac{k-2}{2}} z_g\,
\phi^{(k-1)}\!\left(\frac{x - b z_g}{\sigma\sqrt{z}}\right) dF_Z(z)
< 0,\\[4pt]
\frac{d}{d\sigma}\varphi_{Y_g}^{(k)}(x;b,\sigma)
&=
\sigma^{k-2}\int_0^{+\infty} z^{\frac{k-1}{2}}
\phi^{(k-2)}\!\left(\frac{x - b z_g}{\sigma\sqrt{z}}\right) dF_Z(z)
> 0.
\end{split}
\]
These monotonicity properties imply that $b_1 \ge b_2$ and $c_1 \le c_2$ lead to
\[
\hat{X}_g \succeq_{(k+1)} \hat{Y}_g.
\]
\end{proof}

\begin{remark}Let $Z$ be a bounded, positive random variable. Consider the random variables $\bar{N}_1(\delta_1, \sigma_1)=\delta_1Z+\sigma_1\sqrt{Z}N_1(0, 1)$ and $\bar{N}_2(\delta_2, \sigma_2)=\delta_2Z+\sigma_2\sqrt{Z}N_2(0, 1)$, where $\sigma_1, \sigma_2>0$ are constants, $\delta_1, \delta_2$ are arbitrary real numbers, and $N_1, N_2$ are standard normal random variables independent of $Z$. Our previous analysis establishes that $\bar{N}_1\succeq_{(2)} \bar{N}_2$ holds if and only if $\delta_1\geq \delta_2$ and $\sigma_1\le \sigma_2$. This condition coincides with the usual stochastic dominance ordering for normal random variables $N(\delta_1, \sigma_1)$ and $N(\delta_2, \sigma_2)$. However, in contrast to normal variables, the random variables $\bar{N}_1$ and $\bar{N}_2$ constructed above are generally skewed and need not be symmetric.
    
\end{remark}

\section{Frontier portfolios}

In this section, we investigate solutions to problem (\ref{asiddd}). For ease of exposition, we first fix a positive integer $k \ge 1$ and consider risk measures $\rho: L^k \to \R \cup \{\infty\}$ associated with problem (\ref{asiddd}). Writing $dom(\rho) = \{\eta \in L^k : \rho(\eta) < \infty\}$ for the effective domain of $\rho$, we say that $\rho$ is proper if $\rho(\eta) > -\infty$ for all $\eta \in L^k$ and $dom(\rho) \neq \emptyset$. 

A risk measure on $L^k$ is called a convex risk measure if it is a proper monetary risk measure (that is, it is monotone and cash-invariant) and is convex, meaning
\[
\rho(\lambda \eta_1 + (1-\lambda)\eta_2) \le \lambda \rho(\eta_1) + (1-\lambda)\rho(\eta_2), \quad \forall \lambda \in [0,1],\ \forall \eta_1,\eta_2 \in L^k.
\]
A convex risk measure is said to be coherent if it is positively homogeneous, i.e., $\rho(\lambda \eta) = \lambda \rho(\eta)$ for all $\lambda > 0$. By Corollary 2.3 of \cite{Kaina-Ruschendrof}, any finite convex risk measure on $L^k$ is continuous on $L^k$. In particular, the risk measure $CVaR$ is a coherent risk measure on $L^k$ and is therefore continuous on $L^k$. 

Another key property of $\rho$ for our analysis in this section is its compatibility with second order stochastic dominance. We say that $\rho$ is consistent with second order stochastic dominance (SSD-consistent) if, for any $\eta_1, \eta_2 \in L^k$, the relation $\eta_1 \succeq_{(2)} \eta_2$ implies $\rho(\eta_1) \ge \rho(\eta_2)$. This SSD-consistency of $\rho$ will be needed in the proof of Theorem \ref{efcvarrn} below. We call $\rho$ law-invariant if $\rho(\eta_1) = \rho(\eta_2)$ whenever $\eta_1 \overset{d}{=} \eta_2$. Every finite-valued, law-invariant, convex risk measure is SSD-consistent; see \cite{MaoWang} and the references therein.

\begin{remark}\label{311}
We point out that a solution to problem (\ref{asiddd}) always exists when the risk measure $\rho$ is finite-valued, law-invariant, and convex. To verify this, first note that the mapping $\omega \mapsto \rho(-\omega^T \xi)$ is continuous whenever $\xi \in L_d^k$, since $\rho$ is continuous on $L^k$ as discussed above. Moreover, for any integrable random variable $\eta$, one has $E\eta \succeq_{(2)} \eta$, and thus $\rho(-\eta) \ge \rho(-E\eta)$; see Appendix B of \cite{Svindlandstat} and Corollary 5.1 of \cite{Follmer-Knispel} for analogous arguments. Consequently, on the set $\{\omega \in \R^d : E(-\omega^T X) = r,\ \omega^T e = 1\}$ we have $-r \succeq_{(2)} \omega^T X$, which yields
\[
\rho(-\omega^T X) \ge \rho(r) = r + \rho(0).
\]
\end{remark}

Before presenting the main result of this section, we briefly recall the classical Markowitz mean–variance portfolio optimization problem.  
In the Markowitz framework, portfolio risk is quantified by the variance. The associated optimization problem is
\begin{equation}\label{var1}
\begin{split}
\min_{\omega} Var(-\omega^TX),&\\
E(-\omega^TX)=r,&\\
\omega^T\1=1.&\\
\end{split}    
\end{equation}
A closed-form solution to this problem is well known and can be found in any standard textbook on the capital asset pricing model; see, for instance, page 64 of \cite{chi-fu-huang}. We record the solution here because it will be used in our main results. Let $\mu=EX$ be the mean vector and let $V=Cov(X)$ be the covariance matrix of $X$. For each $r\in \R$, the optimizer of (\ref{var1}) is
\begin{equation}\label{omega0}
\omega^{\star}_r=\omega^{\star}_r(\mu, V)=\frac{1}{d_4}\big[d_2(V^{-1}\1)-d_1(V^{-1}\mu)\big]+\frac{r}{d_4}\big[d_3(V^{-1}\mu)
-d_1(V^{-1}\1)\big],    
\end{equation}
where
\begin{equation*}
d_1=-\1^TV^{-1}\mu, \quad d_2=\mu^TV^{-1}\mu, \quad d_3= \1^TV^{-1}\1, \quad d_4=d_2d_3-d_1^2.  
\end{equation*}
\begin{remark}\label{dfour} Since $V^{-1}$ is positive definite, we obtain $\big(d_1\mu-d_2\1 \big)^TV^{-1}\big(d_1\mu-d_2\1 \big)>0$ and $d_2>0$ (here we assume $\mu=EX\neq 0$). Moreover,
\[
\big(d_1\mu-d_2\1\big)^TV^{-1}\big(d_1\mu-d_2\1\big)=d_2(d_2d_3-d_1^2)=d_2d_4>0,
\]
which implies $d_4>0$ in (\ref{omega0}). 
\end{remark}

In this section, we demonstrate that the optimal portfolio for our mean–risk portfolio optimization problem \eqref{asiddd} can also be written in a form analogous to \eqref{omega0}. To this end, we begin by introducing some notation. For any random vector $\theta$ with mean vector $\mu_{\theta}=E\theta$ and covariance matrix $\Sigma_{\theta}=\mathrm{Cov}(\theta)$, define  
\begin{equation}\label{omega}
\omega^{\star}_{\theta}=\omega^{\star}_{\theta}(\mu_{\theta}, \Sigma_{\theta})=\frac{1}{d_{4\theta}}\bigl[d_{2\theta}(\Sigma_{\theta}^{-1}\1)-d_{1\theta}(\Sigma_{\theta}^{-1}\mu_{\theta})\bigr]+\frac{r}{d_{4\theta}}\bigl[d_{3\theta}(\Sigma_{\theta}^{-1}\mu_{\theta})-d_{1\theta}(\Sigma_{\theta}^{-1}\1)\bigr],    
\end{equation}
where
\begin{equation*}
d_{1\theta}=-\1^T\Sigma_{\theta}^{-1}\mu_{\theta}, \quad d_{2\theta}=\mu_{\theta}^T\Sigma_{\theta}^{-1}\mu_{\theta}, \quad d_{3\theta}= \1^T\Sigma_{\theta}^{-1}\1, \quad d_{4\theta}=d_{2\theta}d_{3\theta}-d_{1\theta}^2.  
\end{equation*}

\begin{remark}
The right-hand side of \eqref{omega} depends only on the mean vector $\mu_{\theta}$ and covariance matrix $\Sigma_{\theta}$ of the random vector $\theta$. Consequently, $\omega^{\star}_{\theta}=\omega^{\star}_{\eta}$ whenever $\theta$ and $\eta$ share the same mean vectors and covariance matrices. We use the notation $\omega^{\star}_{\theta}$ to emphasize that this portfolio is the efficient frontier portfolio, in the mean–variance sense, corresponding to the return vector $\theta$. Note that the formula \eqref{omega} is well defined whenever $\mu_{\theta}\neq 0$ and $\Sigma_{\theta}$ are positive definite (i.e., all eigenvalues are strictly positive), as discussed in Remark \ref{dfour}.
\end{remark}

The next result, Theorem \ref{efcvarrn}, establishes that for problem \eqref{asiddd}, frontier portfolios under any real-valued, law-invariant, convex risk measure always exist and can be written in the form \eqref{omega} when the return vectors are specified by \eqref{one}.

\begin{theorem} \label{efcvarrn}
Fix an integer $k\geq 1$. Let $\rho$ be a finite-valued, law-invariant, convex risk measure on $L^k$. Suppose the return vector is given by \eqref{one} with $Z\in L^k$. Then, for each fixed real number $r$, the solution to problem \eqref{asiddd} is given by $\omega^{\star}_{\theta}$ as in \eqref{omega} with $\mu_{\theta}=\mu+\gamma EZ$ and $\Sigma_{\theta}=\Sigma$.
\end{theorem}

\begin{proof}
Define $D_r=\{\omega \in \mathbb{R}^d: \omega^T\gamma EZ=r,\ \omega^T\1=1\}$. On $D_r$, the minimizer of the quadratic form $\omega^T\Sigma \omega$ is  
$\bar{\omega}:=\omega_{\theta}(\mu+\gamma EZ, \Sigma)$, as in \eqref{omega}. This is frontier portfolio. Then by Lemma \ref{keylem3} below, we have $\bar{\omega}^TX \succeq \omega^TX$ for all other $\omega \in D_r$. The assumptions on $\rho$ in the theorem ensure that $\rho$ is SSD-consistent, and therefore $\rho(-\bar{\omega}^T X)\le \rho(-\omega^T X)$ for all $\omega\in D_r$. This establishes the claim. \qed
\end{proof}

\begin{remark}
Observe that $\mu_{X}=EX=\gamma EZ=\mu_{\theta}$. However, the covariance matrix $ \mathrm{Cov}(X)=\gamma\gamma^T \mathrm{Var}(Z)+A^TAEZ$ differs from $\Sigma_{\theta}=A^TA$. Hence, the optimal solution of \eqref{var1} is not the same as the optimal solution of \eqref{asiddd}. They are nevertheless closely related in that they differ only in the covariance matrices $\Sigma_X$ and $\Sigma_{\theta}$.
\end{remark}

\begin{remark}
The content of Theorem \ref{efcvarrn} can be summarized as follows: the mean–risk efficient frontier portfolios associated with any real-valued, law-invariant, convex risk measure $\rho$ can be obtained by solving a Markowitz mean–variance portfolio optimization problem with a suitably modified return vector $\theta$, as specified in Theorem \ref{efcvarrn}.
\end{remark}

The risk measure $CVaR_{\alpha}$, for each fixed $\alpha \in (0,1)$, is finite-valued, law-invariant, and convex. Consequently, the conclusion of Theorem \ref{efcvarrn} above applies to this risk measure as well. We record this in the following corollary.

\begin{corollary}\label{cvarnew}
Let $X$ be a return vector as in (\ref{one}) with $Z\in L^k$ for some positive integer $k$. Then, for any fixed $\alpha \in (0,1)$, the closed-form solution to the optimization problem
\begin{equation}\label{condvar}
\begin{split}
\min_{\omega}\; &CVaR_{\alpha}(-\omega^T X),\\
\text{s.t.}\;&E(-\omega^T X)=r,\\
&\omega^T \1=1,\\
\end{split}
\end{equation}
is given by $\omega_{\theta}^{\star}$ in (\ref{omega}), where $\mu_{\theta}=EY=\mu+\gamma EZ$ and $\Sigma_{\theta}=A^TA$.
\end{corollary}

\begin{remark}\label{rem-nor}
Theorem \ref{efcvarrn} above can be used to streamline the computation of optimal $CVaR_{\alpha}$ values on certain portfolio domains when returns are normally distributed—a setting analyzed in Theorem 2 of \cite{Rockafellar_R_Tyrrell_And_Uryasev_Stanislav_2000} (see also \cite{Rockafellar_R_Tyrrell_And_Uryasev_Stanislav_2002}). 

To see this, first note that if $Z=1$ in (\ref{one}), then $X$ becomes a Normal random vector. For notational simplicity in what follows, set $\mu=0$ and write $X\sim N_d(\gamma,\Sigma)$. With all other model parameters fixed, $\gamma$ and $\Sigma$ are then fixed, and expression (\ref{omega}) depends only on the target expected return $r$. Thus, for convenience, denote by $\omega_r:=\omega_{\theta}^{\star}$ (for $\theta \overset{d}{=} N_d(\gamma,\Sigma)$ in (\ref{omega})). We can rewrite $\omega_r$ as
\begin{equation*}\label{omegar}
\omega_r = k_1(r)\Sigma^{-1}e + k_2(r)\Sigma^{-1}\gamma,
\end{equation*}
where
\begin{equation*}\label{k1k2}
k_1(r)=\frac{d_{2\theta}}{d_{4\theta}} - r\frac{d_{1\theta}}{d_{4\theta}}, 
\quad
k_2(r)=r\frac{d_{3\theta}}{d_{4\theta}} - \frac{d_{1\theta}}{d_{4\theta}}.  
\end{equation*}
With this notation we have $\omega_r^T N_d \sim N(\omega_r^T\gamma,\sigma_r^2)$, where
\begin{equation*}\label{sigmar}
\sigma_r^2:=\omega_r^T\Sigma\omega_r 
= k_1^2(r)e^T\Sigma^{-1}e 
 + 2k_1(r)k_2(r)e^T\Sigma^{-1}\gamma 
 + k_2^2(r)\gamma^T\Sigma^{-1}\gamma.    
\end{equation*}

From \cite{Panjer2002} (see also equation (2) in \cite{Landsman}), for a Normal random variable $H\sim N(\delta,\sigma^2)$ we have
\begin{equation*}\label{panjar}
CVaR_{\alpha}(H)
= \delta 
+ \Bigg[\frac{1}{\sigma}\,
\frac{\phi\big(\frac{z_{\alpha}-\delta}{\sigma}\big)}
      {1-\varphi\big(\frac{z_{\alpha}-\delta}{\sigma}\big)}\Bigg]\sigma^2,    
\end{equation*}
where $z_{\alpha}$ is the $\alpha$–quantile of the standard Normal distribution $N(0,1)$. For any return level $r$, define
\begin{equation*}
h_{\alpha}(r)
:= r 
+ \Bigg[\frac{1}{\sigma_r}\,
\frac{\phi\big(\frac{z_{\alpha}-r}{\sigma_r}\big)}
      {1-\varphi\big(\frac{z_{\alpha}-r}{\sigma_r}\big)}\Bigg]\sigma_r^2,
\end{equation*}
with $\sigma_r$ as in (\ref{sigmar}).

Now consider the domain
\[
D = \{\omega : \omega^T\gamma \geq \bar{r}\},
\]
for some fixed real number $\bar{r}$. Let $D_{\gamma}:=\{\omega^T\gamma : \omega\in D\}$. Note that $D$ has the property that for any $r_0\in D_{\gamma}$, every portfolio $\omega$ satisfying $\omega^T\gamma=r_0$ belongs to $D$, i.e., $\omega\in D$. On this domain we obtain
\begin{equation*}\label{halpha}
\min_{\omega\in D} CVaR_{\alpha}(\omega^T\theta)
= \min_{r\in D_{\gamma}} h_{\alpha}(r),
\end{equation*}
where $\theta\overset{d}{=}N_d(\gamma,\Sigma)$. The right-hand side of (\ref{halpha}) is simply the minimization of a real-valued function over a subset of the real line, which is significantly simpler than minimizing $F_{\alpha}(\omega,a)$ in (4) of \cite{Rockafellar_R_Tyrrell_And_Uryasev_Stanislav_2000}, as done in Theorem 2 of that paper.
\end{remark}

The proof of Theorem \ref{efcvarrn} relies on  auxiliary results. We begin by introducing some notation. For any real number $r$, define
\begin{equation*}
\mathcal{D}_r=\{\omega \in \R^d: \omega^T(\mu+\gamma EZ)=r, \; \omega^T\1=1\}.
\end{equation*}
Select any two portfolios $\omega_1$ and $\omega_2$ from $\mathcal{D}_r$. We use the following shorthand:
\begin{equation}\label{abc}
\begin{split}
 a_1=\omega_1^T\mu,\;  b_1=\omega_1^T\gamma,\; c_1=\sqrt{\omega_1^T\Sigma \omega_1},\\
 a_2=\omega_2^T\mu,\;  b_2=\omega_2^T\gamma,\; c_2=\sqrt{\omega_2^T\Sigma \omega_2},\\
\end{split}
\end{equation}
for any pair of portfolios $\omega_1, \omega_2\in \mathcal{D}_r$. Define
\begin{equation*}
\begin{split}
 \eta_1=&a_1+b_1Z+c_1\sqrt{Z}N,\\
 \eta_2=&a_2+b_2 Z+c_2\sqrt{Z}N,\\
\end{split}    
\end{equation*}
where the coefficients $a_i, b_i, c_i$ are specified as in (\ref{abc}). We aim to determine conditions that guarantee 
\[
\eta_1\succeq \eta_2.
\]

\begin{remark}\label{rem3.10} Consider any two portfolios $\omega_1$ and $\omega_2$ in the set $\mathcal{D}_r$, and define $\omega_0 = \omega_2 - \omega_1$. Then it follows immediately that $\omega_0^T(\mu + \gamma EZ) = 0$ and $\omega_0^T\1=0$.  Define
\[
\mathcal{H} = \{\omega \in \R^d : \omega^T(\mu + \gamma EZ) = 0, \; \omega^T\1=0\}.
\]
 With this definition, we can write
\[
\mathcal{D}_r = \{\omega_1 + \omega_0 : \omega_0 \in \mathcal{H}\},
\]
for any fixed choice of $\omega_1 \in \mathcal{D}_r$.

Given any two portfolios $\omega_1, \omega_2 \in \mathcal{D}_r$, let $\eta_1 = \omega_1^T X$ and $\eta_2 = \omega_2^T X$. With $\omega_0 = \omega_2 - \omega_1$, we can write  $\eta_2 = \eta_1 + \omega_0^T X$. Observe that
\[
\eta_1 \overset{d}{=} \omega_1^T\mu + \omega_1^T\gamma Z + \sqrt{\omega_1^T \Sigma \omega_1}\,\sqrt{Z}N,
\]
and
\[
\eta_2 \overset{d}{=} (\omega_1^T\mu + \omega_0^T\mu) + (\omega_1^T\gamma + \omega_0^T\gamma)Z + \sqrt{(\omega_1 + \omega_0)^T \Sigma (\omega_1 + \omega_0)}\,\sqrt{Z}N.
\]

From the condition $\omega_0^T(\mu + \gamma EZ) = 0$, we deduce that, assuming both components are nonzero, either
\[
\omega_0^T\mu > 0,\quad \omega_0^T\gamma < 0,
\]
or
\[
\omega_0^T\mu < 0,\quad \omega_0^T\gamma > 0.
\]
Thus, when comparing $\eta_1$ and $\eta_2$, we are in one of the following two cases:
\[
\omega_1^T\mu > \omega_1^T\mu + \omega_0^T\mu,\quad \omega_1^T\gamma < \omega_1^T\gamma + \omega_0^T\gamma,
\]
or
\[
\omega_1^T\mu < \omega_1^T\mu + \omega_0^T\mu,\quad \omega_1^T\gamma > \omega_1^T\gamma + \omega_0^T\gamma,
\]
under the assumption that $\omega_0^T\gamma \neq 0$. We define 
\[
\bar{\eta}_2=\omega_1^T\mu + \omega_1^T\gamma Z + \sqrt{(\omega_1 + \omega_0)^T \Sigma (\omega_1 + \omega_0)}\,\sqrt{Z}N,
\]
and we observe that $\eta_2=\bar{\eta}_2+\omega_0^T\mu+\omega_0^T\gamma EZ$.
\end{remark}

\begin{remark}\label{rem3.11} Consider the setting in Remark \ref{rem3.10} above. We  have $\omega_2=\omega_1+\omega_0$ and $c_2=\sqrt{(\omega_1+\omega_0)^T\Sigma (\omega_1+\omega_0)}=\sqrt{\omega_1^T\Sigma \omega_1+2\omega_1^T\Sigma \omega_0+\omega_0^T\Sigma \omega_0}$. If $\omega_1$ is a frontier portfolio, then it takes the form $\omega_1=k_1\Sigma^{-1}(\mu+\gamma EZ)+k_2\Sigma^{-1}\1$ for some real numbers $k_1, k_2$ (see (\ref{omega0}) for this). Since $\omega_0\in \mathcal{H}$ we have $\omega_0^T(\mu+\gamma EZ)=0$ and $\omega_0^T\1=0$. From these one can conclude that $\omega_0^T\Sigma \omega_1=0$ as long as $\omega_1$ is a frontier portfolio.
\end{remark}

\begin{lemma}\label{lbound}Let $\omega_1$ and $\omega_2$ be any two portfolios in $\mathcal{D}_r$ for a fixed real number $r$, and let $\eta_1, \eta_2, \bar{\eta}_2$ be defined as in Remark \ref{rem3.10}. Let $\omega_0=\omega_2-\omega_1$ and assume that
\[
2\omega_1^T\Sigma \omega_0+\omega_0^T\Sigma \omega_0>0.
\]
Then, for all $x\in \R$, we have  
\[
F^{(2)}_{\bar{\eta}_2}(x)-F^{(2)}_{\eta_1}(x)\geq (c_2-c_1)\int_{0}^{\infty}\sqrt{z}\,\phi^{(0)}\!\Big(\frac{x-a_1-b_1z}{c_1\sqrt{z}}\Big)dF_Z(z),
\]
where $c_2=\sqrt{(\omega_1+\omega_0)^T\Sigma (\omega_1+\omega_0)}$, $c_1=\sqrt{\omega_1^T\Sigma \omega_1}$, $a_1=\omega_1^T\mu$, and $b_1=\omega_1^T\gamma$.
\end{lemma}
\begin{proof} Note that $F^{(2)}_{\eta_1}(x)=E(x-\eta_1)^+$ and $F^{(2)}_{\bar{\eta}_2}(x)=E(x-\bar{\eta}_2)^+$. We have
\[
F^{(2)}_{\bar{\eta}_2}(x)-F^{(2)}_{\eta_1}(x)=\int_{0}^{\infty}[c_2\sqrt{z}\phi^{(2)}(\frac{x-a_1-b_1z}{c_2\sqrt{z}})-c_1\sqrt{z}\phi^{(2)}(\frac{x-a_1-b_1z}{c_1\sqrt{z}})]dF_Z(z)
\]
We define the function $Q_z(c)=:c\sqrt{z}\phi^{(2)}(\frac{x-a_1-b_1z}{c\sqrt{z}})$ for each fixed $z, x, a_1, b_1$. We have
\[
Q_z'(c)=\sqrt{z}[\phi^{(2)}(\frac{x-a_1-b_1z}{c\sqrt{z}})-\frac{x-a_1-b_1z}{c\sqrt{z}}\phi^{(1)}(\frac{x-a_1-b_1z}{c\sqrt{z}})]=\sqrt{z}\phi^{(0)}(\frac{x-a_1-b_1z}{c\sqrt{z}}).
\]
Here the second equality follows from the relation $\phi^{(2)}(x)-x\phi(x)=\phi^{(0)}(x)$ due to   (\ref{phinn}). By the middle value Theorem, we have $Q_z(c_2)-Q_z(c_1)=\sqrt{z}\phi^{(0)}(\frac{x-a_1-b_1z}{\bar{c}\sqrt{z}})$ for some $c_1<\bar{c}<c_2$. Now if $x-a_1-b_1z\geq 0$, then $\frac{x-a_1-b_1z}{\bar{c}\sqrt{z}}\le \frac{x-a_1-b_1z}{c_1\sqrt{z}}$ and therefore $\phi^{(0)}(\frac{x-a_1-b_1z}{\bar{c}\sqrt{z}})\geq \phi^{(0)}(\frac{x-a_1-b_1z}{c_1\sqrt{z}})$ (recall here that $\phi^{(0)}(x)=\frac{1}{\sqrt{2\pi}}e^{-\frac{x^2}{2}}$). If $x-a_1-b_1z<0$, then 
$|\frac{x-a_1-b_1z}{\bar{c}\sqrt{z}}|\le |\frac{x-a_1-b_1z}{c_1\sqrt{z}}|$ and therefore we still have $\phi^{(0)}(\frac{x-a_1-b_1z}{\bar{c}\sqrt{z}})\geq \phi^{(0)}(\frac{x-a_1-b_1z}{c_1\sqrt{z}})$. From these we conclude that
\[
F^{(2)}_{\bar{\eta}_2}(x)-F^{(2)}_{\eta_1}(x)=\int_{0}^{\infty}[Q_z(c_2)-Q_z(c_1)]f(z)dz\geq (c_2-c_1)\int_{0}^{\infty}\sqrt{z}\phi^{(0)}(\frac{x-a_1-b_1z}{c_1\sqrt{z}})dF_Z(z).
\]
    
\end{proof}

Recall that our objective is to establish $\eta_1 \succ \eta_2$, which is equivalent to proving that $F_{\eta_2}^{(2)}(x)\geq F_{\eta_1}^{(2)}(x)$ for every real $x$, with the inequality being strict for at least one value of $x$. Observe that
\[
F_{\eta_2}^{(2)}(x)-F_{\eta_1}^{(2)}(x) = \bigl(F_{\eta_2}^{(2)}(x)- F_{\bar{\eta}_2}^{(2)}(x)\bigr) + \bigl(F_{\bar{\eta}_2}^{(2)}(x)-F_{\eta_1}^{(2)}(x)\bigr),
\]
and since $\bigl(F_{\bar{\eta}_2}^{(1)}(x)-F_{\eta_1}^{(1)}(x)\bigr)$ admits a lower bound as given in Lemma \ref{lbound}, it is enough to verify the assertion stated in the following lemma.

\begin{lemma} \label{keylem3} Let $\omega_1$ and $\omega_2$ be any two portfolios in $\mathcal{D}_r$ for a fixed real number $r$, and let $\eta_1, \eta_2, \bar{\eta}_2$ be defined as in Remark \ref{rem3.10}. Let $\omega_0=\omega_2-\omega_1$. Assume that $\omega_1$ is a frontier portfolio as in (\ref{omega0}). Then for every $x \in \R$ we have
\begin{equation}\label{keyineq}
F^{(2)}_{\eta_2}(x)-F^{(2)}_{\bar{\eta}_2}(x)\geq -(c_2-c_1)\int_{0}^{\infty}\sqrt{z}\,\phi^{(0)}\!\left(\frac{x-a_1-b_1z}{c_1\sqrt{z}}\right)\,dF_Z(z),
\end{equation}
where the constants $a_1, b_1, c_1, c_2$ are as specified in Lemma \ref{lbound}. 
\end{lemma}

\begin{proof} First note that, since $\omega_1$ lies on the efficient frontier, Remark \ref{rem3.11} implies
\[
c_2=\sqrt{\omega_2^T\Sigma \omega_2}=\sqrt{\omega_1^T\Sigma \omega_1+\omega_0^T\Sigma \omega_0}>c_1
\]
whenever $\omega_0\in \mathcal{H}$ is nonzero. Next, we can express the difference $F^{(2)}_{\eta_2}(x)-F^{(2)}_{\bar{\eta}_2}(x)$ as
\[
F^{(2)}_{\eta_2}(x)-F^{(2)}_{\bar{\eta}_2}(x)=\int_0^{\infty}c_2\sqrt{z}\Big[\phi^{(2)}\!\Big(\frac{x-a_1-b_1z}{c_2\sqrt{z}}-\frac{\delta_z(\omega_0)}{c_2\sqrt{z}}\Big)-\phi^{(2)}\!\Big(\frac{x-a_1-b_1z}{c_2\sqrt{z}}\Big)\Big]\,dF_{Z}(z),
\]
where we define $\delta_z(\omega_0):=\omega_0^T\mu+\omega_0^T\gamma z$. By the convexity of $\phi^{(2)}(\cdot)$, we obtain
\begin{equation*}
\begin{split}
F^{(2)}_{\eta_2}(x)-F^{(2)}_{\bar{\eta}_2}(x)\geq  & -\int_0^{\infty}\delta_z(\omega_0)\,\phi^{(1)}\Big(\frac{x-a_1-b_1z}{c_2\sqrt{z}}\Big)\,dF_Z(z)\\
=&-\omega_0^T\mu\int_0^{\infty}\phi^{(1)}\Big(\frac{x-a_1-b_1z}{c_2\sqrt{z}}\Big)\,dF_Z(z)\\
-&\omega_0^T\gamma\int_0^{\infty}z\,\phi^{(1)}\Big(\frac{x-a_1-b_1z}{c_2\sqrt{z}}\Big)\,dF_Z(z).\\
\end{split}
\end{equation*}

\emph{Step 1:} 
Define
\begin{equation*}
\begin{split}
L(x; \omega_0)=&-\int_0^{\infty}\delta_z(\omega_0)\,\phi^{(1)}\!\left(\frac{x-a_1-b_1z}{c_2\sqrt{z}}\right)\,dF_Z(z)\\
&\quad+(c_2-c_1)\int_{0}^{\infty}\sqrt{z}\,\phi^{(0)}\!\left(\frac{x-a_1-b_1z}{c_1\sqrt{z}}\right)\,dF_Z(z).
\end{split}
\end{equation*}
Observe that to establish (\ref{keyineq}), it is enough to show that $L(x; \omega_0)\geq 0$ for all $x$ and $\omega_0$.
To this purpose, using the identity $\omega_0^T\mu+\omega_0^T\gamma EZ=0$, we express $L(x; \omega_0)$ in the following two equivalent forms:
\begin{equation*}
 \begin{split}
 L_1(x; \omega_0)=&\;\omega_0^T\gamma\, J(x, c_2)
 +(c_2-c_1)\,Q(x),\\  
L_2(x; \omega_0)=&\;\omega_0^T\mu\, H(x, c_2)
 +(c_2-c_1)\,Q(x),
 \end{split}   
\end{equation*}
where
\begin{equation*}
\begin{split}
J(x, \theta)=&\int_0^{\infty}(EZ-z)\,\phi^{(1)}\!\left(\frac{x-a_1-b_1z}{\theta\sqrt{z}}\right)dF_Z(z),\\ 
H(x, \theta)=&\int_0^{\infty}\left(\frac{z}{EZ}-1\right)\phi^{(1)}\!\left(\frac{x-a_1-b_1z}{\theta \sqrt{z}}\right)dF_Z(z),\\
Q(x)=&\int_{0}^{\infty}\sqrt{z}\,\phi^{(0)}\!\left(\frac{x-a_1-b_1z}{c_1\sqrt{z}}\right)dF_Z(z).
\end{split}    
\end{equation*}

\emph{Step 2:} In this step we prove that the mapping $\omega_0 \mapsto L(x; \omega_0)$ is bounded from below on the set $\mathcal{H}$ introduced in Remark \ref{rem3.10}. To this end, we work with the equivalent representation $L_1(x; \omega_0)$ of $L$. Note first that $(c_2 - c_1)Q(x)$ is strictly positive for every $x$. Moreover, $L_1(x; \omega_0)$ is a bounded function on any bounded subset of $\mathcal{H}$. The only potentially negative contribution to $L_1$ stems from the term $\omega_0^T \gamma J(x, c_2)$. Thus, to establish that $L_1$ is bounded from below for each fixed $x$, it is enough to verify that
\[
\frac{\omega_0^T \gamma J(x, c_2)}{(c_2 - c_1)Q(x)} \to 0,
\]
whenever $\omega_0^T \Sigma \omega_0 \to \infty$. Since $Q(x)$ is a fixed, finite, positive constant for each fixed $x$, and
\[
\lim_{c_2 \to \infty} J(x, c_2)
= \int_0^\infty (EZ - z)\,\lim_{c_2 \to \infty} \phi^{(1)}\!\Big(\frac{x - a_1 - b_1 z}{c_2 \sqrt{z}}\Big)\,dF_Z
= \tfrac{1}{2}\int_0^\infty (EZ - z)\,dF_Z(z)
= 0,
\]
it suffices to show that $\limsup_{c_2 \to \infty}\bigl|\frac{\omega_0^T \gamma}{c_2 - c_1}\bigr|$ is finite. For this, observe that
\[
c_2 - c_1
= \sqrt{\omega_1^T \Sigma \omega_1 + \omega_0^T \Sigma \omega_0}\;-\;\sqrt{\omega_1^T \Sigma \omega_1},
\]
and, by the Cauchy–Schwarz inequality,
\[
|\omega_0^T \gamma|
= |\omega_0^T A A^{-1} \gamma|
\le \sqrt{\omega_0^T \Sigma \omega_0}\,\sqrt{\gamma^T \Sigma^{-1} \gamma}.
\]
This implies
\[
\limsup_{c_2 \to \infty}\Bigl|\frac{\omega_0^T \gamma}{c_2 - c_1}\Bigr|
\le \sqrt{\gamma^T \Sigma^{-1} \gamma}.
\]
We conclude that $L$ is bounded from below for every fixed $x$.

\emph{Step 3:} Note that $\frac{dc_2}{d\omega_0}=\frac{1}{c_2}\Sigma \omega_0$. Recall that our objective is to prove that, for each fixed $x$, we have $L(x,\omega_0)\geq 0$ for all $\omega_0$. Consider the domain
\[
\mathcal{D}_{\omega_0}=\{\omega_0:\omega_0^T\Sigma\omega_0\geq 0\}.
\]
Its boundary is 
\[
\partial\mathcal{D}_{\omega_0}=\{\omega_0:\omega_0^T\Sigma\omega_0=0\}=\{0\},
\]
and its interior is
\[
\mathrm{int}(\mathcal{D}_{\omega_0})=\{\omega_0:\omega_0^T\Sigma\omega_0>0\}.
\]
The mapping $\omega_0\mapsto L(x;\omega_0)$ is clearly continuously differentiable on $\mathrm{int}(\mathcal{D}_{\omega_0})$, and for each fixed $x$ we have $L(x;0)=0$. Our aim is to show that, for each fixed $x$, the function $\omega_0\mapsto L(x;\omega_0)$ does not attain a minimum in $\mathrm{int}(\mathcal{D}_{\omega_0})$. To this end, we examine the first-order condition. We obtain
\begin{equation*}\label{linear1}
\begin{split}
\frac{dL_1}{d\omega_0}=&\,J(x,c_2)\gamma+\frac{1}{c_2}\bigl[\omega_0^T\gamma \,\frac{dJ}{d\theta}\big|_{\theta=c_2}+Q(x)\bigr]\Sigma\omega_0=0,\\[4pt]
\frac{dL_2}{d\omega_0}=&\,H(x,c_2)\mu+\frac{1}{c_2}\bigl[\omega_0^T\mu \,\frac{dH}{d\theta}\big|_{\theta=c_2}+Q(x)\bigr]\Sigma\omega_0=0.\\
\end{split}    
\end{equation*}
This system is equivalent to
\begin{equation*}\label{linear2}
\begin{split}
J(x,c_2)\Sigma^{-1}\gamma+\frac{1}{c_2}\bigl[\omega_0^T\gamma \,\frac{dJ}{d\theta}\big|_{\theta=c_2}+Q(x)\bigr]\omega_0=&\,0,\\[4pt]
H(x,c_2)\Sigma^{-1}\mu+\frac{1}{c_2}\bigl[\omega_0^T\mu \,\frac{dH}{d\theta}\big|_{\theta=c_2}+Q(x)\bigr]\omega_0=&\,0.\\
\end{split}    
\end{equation*}
Next, observe that
\begin{equation*}
\begin{split}
(\mu+\gamma EZ)^T\Sigma^{-1}(\mu+\gamma EZ)
=&\,[\mu^T\Sigma^{-1}\mu+\mu^T\Sigma^{-1}\gamma EZ]\\
&+[\mu^T\Sigma^{-1}\gamma+\gamma^T\Sigma^{-1}\gamma EZ]EZ>0. 
\end{split}
\end{equation*}
Consequently, either $\mu^T\Sigma^{-1}\mu+\mu^T\Sigma^{-1}\gamma EZ>0$ or $\mu^T\Sigma^{-1}\gamma+\gamma^T\Sigma^{-1}\gamma EZ>0$. 

If 
\[
\mu^T\Sigma^{-1}\mu+\mu^T\Sigma^{-1}\gamma EZ=\mu^T\Sigma^{-1}(\mu+\gamma EZ)>0,
\]
then, using $\omega_0^T(\mu+\gamma EZ)=0$, we deduce that $\mu^T\Sigma^{-1}$ is linearly independent of $\omega_0$. Likewise, if
\[
\mu^T\Sigma^{-1}\gamma+\gamma^T\Sigma^{-1}\gamma EZ=(\mu^T+\gamma^TEZ)\Sigma^{-1}\gamma>0,
\]
then, since again $\omega_0^T(\mu+\gamma EZ)=0$, we infer that $\omega_0$ is linearly independent of $\Sigma^{-1}\gamma$ (note that here $J(x,c_2)\neq 0$ and $H(x,c_2)\neq 0$). Hence, in either case, one of the equalities in (\ref{linear2}) cannot be satisfied. Therefore, $L$ cannot have a minimizer in $\mathrm{int}(\mathcal{D}_{\omega_0})$. It must then attain its minimum on the boundary $\partial\mathcal{D}_{\omega_0}=\{0\}$. Since $L(x,0)=0$, we conclude that $L$ is nonnegative for all $x$ and $\omega_0$.
\end{proof}

As an application of the above results, we now present the following example.
 
\begin{example}\label{ex1-1} Let $R$ denote the return vector of $d$ risky assets and suppose that 
\[
R\sim GH_d(\lambda, \alpha, \beta, \delta, \mu, \Sigma),
\]
that is, $R$ follows a multivariate generalized hyperbolic distribution. For the definition of this distribution and the admissible parameter ranges, we refer to Chapter 2 of \cite{hammerstein2010generalized}. As stated in the first paragraph on page 78 of \cite{hammerstein2010generalized}, $R$ admits the NMVM representation
\[
R\overset{d}{=}\mu+Z\Sigma \beta+\sqrt{Z}AN_d,
\]
where $A=\Sigma^{-\frac{1}{2}}$. Consider the optimization problem (\ref{condvar}) with $X=R$,
for an arbitrary but fixed real number $r$ and any fixed $\alpha \in (0, 1)$. By Theorem \ref{efcvarrn}, this problem has a closed-form solution, given by $\omega_{\theta}(\mu+\Sigma \beta EZ, \; \Sigma)$ as in (\ref{omega}). Observe that the mixing variable $Z$ satisfies $Z\sim GIG(\lambda, \delta, \sqrt{\alpha^2-\beta^T\Sigma \beta})$. We further assume that the parameters $\lambda, \alpha, \beta, \delta, \mu, \Sigma$ are chosen so that $Z\in L^k$ for some integer $k\geq 1$; see page 11 of \cite{hammerstein2010generalized} for the moment formulas of GIG distributions. This observation leads to a substantial simplification of portfolio selection problems under GH models. Indeed, Corollary 2.12 of \cite{hammerstein2010generalized} implies that for any $\omega$, one has $\omega^TR\sim GH(\lambda, \hat{\alpha}, \hat{\beta}, \hat{\delta}, \hat{\mu})$, where the parameters $\hat{\alpha}, \hat{\beta}, \hat{\delta}, \hat{\mu}$ are given by rather involved expressions. Having the efficient frontier portfolios in closed form allows one to use the approach from Remark \ref{rem-nor} to compute optimal portfolios under the widely used risk measure $CVaR_{\alpha}$ in the GH setting.
\end{example}

\begin{remark} We note that the conclusion of Example \ref{ex1-1} above continues to hold for any risk measure that is both law-invariant and SSD-consistent. As an illustration, consider the following family of risk measures, which satisfy these requirements:
\begin{equation}\label{last1}
\hat{\rho}(X)=\sup_{\mathcal{P} \in \mathcal{M}}\int_{(0, 1]}CVaR_{\beta}(X)\mathcal{P}(d\beta),    \end{equation}
where $\mathcal{M}$ denotes an arbitrary subset of $\mathcal{M}_1((0, 1])$, the collection of probability measures on $(0, 1]$. For additional details on this class of risk measures, we refer to Section 5.1 and the preceding sections of \cite{Follmer-Knispel}. Furthermore, any finite-valued law-invariant coherent risk measure admits a representation of the above form; see, for instance, the paragraph preceding Corollary 5.1 in \cite{Follmer-Knispel}.
\end{remark}

Next, consider the optimization problem
\begin{equation}\label{aasi}
\begin{split}
\min_{\omega}\; &\rho(-\omega^TX),\\
s.t.\; &E(-\omega^TX )=r,\\
\end{split}    
\end{equation}
for an arbitrary real number $r$. Unlike problem (\ref{asiddd}), in (\ref{aasi}) we no longer impose the normalization constraint $\omega^T\1=1$. The solution to (\ref{aasi}) is related to the following optimization problem:
\begin{equation*}\label{sigma-m}
\begin{split}
\min_{\omega}\; &\omega^T\Sigma \omega,\\
s.t.\; &E(-\omega^TX )=r,\\
\end{split}    
\end{equation*}
whose optimizer is known to be of the form
\[
\bar{\omega}_r=\frac{-r}{(\mu+\gamma EZ)^T\Sigma^{-1}(\mu+\gamma EZ)}\Sigma^{-1}(\mu+\gamma EZ).
\]
We refer to this optimizer as a frontier portfolio. Set $\bar{\mathcal{H}_0}=\{\omega \in \R^d: \omega^T(\mu+\gamma EZ)=0\}$. Then for any two portfolios $\omega_1$ and $\omega_2$ satisfying $E(-\omega^T X)=r$, we have $\bar{\omega}_0=\omega_2-\omega_1\in \bar{\mathcal{H}}_0$. Furthermore, observe that $\bar{\omega}_0^T\Sigma \omega_r=0$. Repeating the reasoning used in the proof of Lemma \ref{keylem3}, we obtain the following corollary.

\begin{corollary} Let $\omega_1$ and $\omega_2$ be any two portfolios in $\bar{E}_r := \{\omega \in \R^d : E(-\omega^T X) = r\}$ for some fixed real number $r$. If $\omega_1$ is the frontier portfolio $\bar{\omega}_r$ defined above, then  $\omega_1^T X\succeq_{(2)} \omega_2^T X$. In addition, the optimizer of problem (\ref{aasi}) is precisely $\bar{\omega}_r$.  
\end{corollary}
\begin{proof} The claim follows from arguments analogous to those in the proof of Lemma \ref{keylem3}.
\end{proof}

\section{CAPM with NMVM-distributed returns}

In this section, we analyze the capital asset pricing model (CAPM) within our proposed framework. The classical CAPM relies on the assumption that asset returns are normally distributed. Under this normality assumption, variance naturally serves as the relevant risk measure, and the mean–variance optimization problem becomes a quadratic program that admits an explicit closed-form solution. However, when asset returns deviate from normality, the traditional CAPM framework loses its rigorous theoretical underpinning.

To address this, we formulate a CAPM under a broader class of NMVM distributions. In our setting, we posit that all asset returns are driven by a common mixing distribution that is almost surely the same across assets. This mixing distribution may be any random variable belonging to the space $L^1$.  Under this assumption, we derive the corresponding CAPM. 

By loosening the distributional assumptions on asset returns in the CAPM context, we substantially extend the model’s scope. As discussed in the introduction, NMVM models reproduce the key empirical characteristics of return distributions, so a CAPM based on this framework is more realistic than one derived under normality. We now explain how to build the CAPM framework within NMVM models.

We fix a mixing distribution $Z$ and define 
\[
\mathcal{D}(Z)=\{a+bZ+c\sqrt{Z}N: a, b\in \R, c\geq 0\},
\]
where $N$ is a standard Normal random variable independent from $Z$. For any two elements $\eta_1=a_1+b_1Z+c_1\sqrt{Z}N_1$ and $\eta_2=a_2+b_2Z+c_2\sqrt{Z}N_2$,  the addition is defined by $\eta_1+\eta_2\overset{d}{=}(a_1+a_2)+(b_1+b_2)Z+\sqrt{c_1^2+2c_1c_2\sigma_{12}+c_2^2}\sqrt{Z}N$, where $\sigma_{12}$ is the correlation between $N_1$ and $N_2$. For any real number $\alpha$ and $\eta=a+bZ+c\sqrt{Z}N\in \mathcal{D}(Z)$ the multiplication $\alpha \eta\overset{d}{=}\alpha a+\alpha bZ+c|\alpha|N $ is in $\mathcal{D}(Z)$. Therefore $\mathcal{D}(Z)$ is a linear space. We assume $Z\in L^k$ for some positive integer $k\geq 1$. Fix any coherent risk measure $\rho_0$ on $L^k$. On $\mathcal{D}(Z)$ we define the following risk measure
\begin{equation}\label{rhotilde}
 \rho(\eta)=a+bEZ+c\rho_0(\sqrt{Z}N).   
\end{equation}
\begin{lemma} The measure $\rho$ is a coherent risk measure on $\mathcal{D}(Z)$.
\end{lemma}
\begin{proof} Note that for two elements $\eta_1=a_1+b_1Z+c_1\sqrt{Z}N_1$ and $\eta_2=a_2+b_2Z+c_2\sqrt{Z}N_1$ in $\mathcal{D}(Z)$, a necessary condition for $\eta_1\geq \eta_2$ is that $c_1=c_2$ and 
$a_1+b_1EZ\geq a_2+b_2EZ$. Hence, the risk measure in (\ref{rhotilde}) is monotone. It is also evidently translation invariant and positively homogeneous. To verify subadditivity, note that
\begin{equation*}
\begin{split}
\rho(\eta_1+\eta_2)=&(a_1+a_2)+(b_1+b_2)EZ+\sqrt{c_1^2+2c_1c_2\sigma_{12}+c_2^2}\,\rho_0(\sqrt{Z}N)\\
=&(a_1+a_2)+(b_1+b_2)EZ+(c_1+c_2)\rho_0(\sqrt{Z}N)=\rho(\eta_1)+\rho(\eta_2),
\end{split}
\end{equation*}
since $\sqrt{c_1^2+2c_1c_2\sigma_{12}+c_2^2}\le c_1+c_2$. Consequently, $\rho$ given in (\ref{rhotilde}) is a coherent risk measure.
\end{proof}

\textbf{Tangent portfolio:}  The tangent portfolio is designed to achieve the best possible balance between risk and return. When a risk-free asset is available, it becomes the single optimal risky portfolio for every investor, no matter how risk-averse they are. Among all feasible portfolios composed solely of risky assets, the tangent portfolio attains the maximum Sharpe Ratio. Within the framework of the Capital Asset Pricing Model (CAPM), the tangent portfolio acquires an even more prominent interpretation: it corresponds to the Market Portfolio. 

To derive the tangent portfolio in our setting, we consider a scenario where investors can allocate their wealth across $d$ risky assets, whose return vector is specified in (\ref{one}), and a risk-free asset with return $r_f$. For a portfolio $\omega \in \R^d$, the resulting portfolio return is
\[
R_{\omega}=\omega^TX+(1-\omega^T \mathbf{1} )r_f.
\]
Under the risk measure $\rho$ defined in (\ref{rhotilde}), the associated risk is
\begin{equation}\label{633}
\begin{split}
\rho(R_{\omega})=&\omega^T(\mu+\gamma EZ)+\sqrt{\omega^T\Sigma \omega}\rho(\sqrt{Z}N)+(1-\omega^T \mathbf{1} )r_f\\
=&r_f+\omega^T(\mu-\1r_f+\gamma EZ)+\sqrt{\omega^T\Sigma \omega}\rho_0(\sqrt{Z}N).
\end{split}
\end{equation}
In the next result, we identify the corresponding tangent portfolio.

\begin{proposition}\label{4.33}
The market portfolio in this setting is
\[
\omega_m=\frac{1}{\mathbf{1}^T\Sigma^{-1}(\mu-\mathbf{1} r_f+\gamma EZ)}\Sigma^{-1}(\mu-\mathbf{1} r_f+\gamma EZ).
\]
The associated market return can be written as
\[
R_m=\omega_m^TX=u_{m}+\beta_mZ+\sigma_m \sqrt{Z}N,
\]
where
\[
u_m=\omega_m^T\mu,\; \beta_m=\omega_m^T\gamma, \; \sigma_m=\sqrt{\omega_m^T\Sigma \omega_m}.
\]
Moreover,
\[
\rho(R_m)=u_m+\gamma_mEZ+\sigma_m \rho_0.
\]
\end{proposition}
\begin{proof}
We consider the maximization of the Sharpe ratio
\[
\frac{ER_{\omega}-r_f}{\rho(R_{\omega})}=\frac{\omega^T(\mu-\1 r_f+\gamma EZ)}{\rho(R_{\omega})},
\]
over all portfolios satisfying $\omega^T(\mu-\1 r_f+\gamma EZ)>0$. This condition guarantees that $\rho(R_{\omega})>0$. First, note that from (\ref{633}) we have
\[
\frac{d\rho(R_{\omega})}{d\omega}=(\mu-\1 r_f+\gamma EZ)+\frac{\rho_0}{\sqrt{\omega^T\Sigma \omega}}\Sigma \omega. 
\]
Imposing the first-order condition yields that the optimal portfolio $\omega^{\star}$ is of the form
\[
\omega^{\star}=L(\omega^{\star})\Sigma^{-1}(\mu-\1 r_f+\gamma EZ),
\]
with
\[
L(\omega^{\star})=\frac{\rho(R_{\omega^{\star}})-(\omega^{\star})^T(\mu-\1r_f+\gamma EZ)}{(\omega^{\star})^T(\mu-\1r_f+\gamma EZ)}\frac{\sqrt{(\omega^{\star})^T\Sigma \omega^{\star}}}{\rho_0}.
\]
Normalizing by the total weight, we obtain
\[
\omega_m=\frac{\omega^{\star}}{ \1^T\omega^{\star}},
\]
which coincides with the expression stated in the proposition. The remaining assertions follow directly from the definitions.
\end{proof}

In what follows, we explore several important implications of the results derived in this paper. We first introduce the following risk measure:
\begin{equation}\label{new-risk}
\bar{\sigma}(\eta)=\frac{\rho(\eta)-E\eta}{\rho(\sqrt{Z}N)}=\frac{\rho(\eta)-E\eta}{\rho_0},
\end{equation}
for an arbitrary NMVM random variable $\eta=a+b\eta+c\sqrt{Z}N$. We clearly have $\bar{\sigma}(\eta)=c$. The risk measure $\bar{\sigma}(\cdot)$ defined in (\ref{new-risk}) is not coherent. Rather, it can be viewed as a risk measure that is analogous to the conventional standard deviation. With respect to this risk measure, we immediately obtain $\bar{\sigma}(R_{\omega})=\sqrt{\omega^T\Sigma \omega}$ and $\bar{\sigma}^2(R_{\omega})=\omega^T\Sigma \omega$ (see also (\ref{633})). Note that this measure corresponds to the deviation measure studied in \cite{Tyrrell}.

\textbf{CAPM based on the mean–$\bar{\sigma}$ optimization criterion:} 
The CAPM, grounded in the mean–variance framework, has long served as a foundation of modern financial economics, as it quantifies the trade-off between risk and return. CAPM demonstrates that systematic risk cannot be diversified away, while idiosyncratic risk can be fully eliminated by holding a well-diversified portfolio. It also offers a standard benchmark for determining the cost of capital. Using the CAPM formula, investors and fund managers can assess the performance of their investment in an objective manner: one can compute the “expected” return of a fund given its level of risk (beta) and the performance of the overall market.The next theorem derives the version of the CAPM appropriate for our framework. We refer to this specification as the skewness-induced CAPM (SI-CAPM).

\begin{theorem}\label{capm-new} For any return of the form $R = a + bZ + c\sqrt{Z}N_r$, we have
\begin{equation}\label{skew-capm}
ER - r_f = \sigma_{rm}\frac{\rho(R) - ER}{\rho(R_m) - ER_m}(ER_m - r_f),
\end{equation}
where $R_m = u_m + \beta_m Z + \sigma_m \sqrt{Z} N_m$ denotes the market return introduced in Proposition \ref{4.33}, and $\sigma_{rm}$ represents the correlation between the standard Normal random variable $N_r$ and the standard Normal variable $N_m$. Under the condition that $Z$ is square integrable, the correlation $\sigma_{rm}$ is given by
\begin{equation}\label{sigrm}
\sigma_{rm}=\frac{Cov(R, R_m)-b\beta_m Var(Z)}{c\sigma_mEZ},
\end{equation}
and the relation (\ref{skew-capm}) can  be written as
\begin{equation}\label{SI-capm}
 ER-r_f=\beta_{SI}(ER_m-r_f),   
\end{equation}
where 
\[
\beta_{SI}=\frac{Cov(R, R_m)-b\beta_mVar(Z)}{Var(R_m)-\beta_m^2Var(Z)},
\]
is the skewness induced beta.
\end{theorem}
\begin{proof} We proceed along the lines of the standard CAPM proof. Consider a risky asset $i$ with return $R_i = a_i + b_i Z + c_i \sqrt{Z} N_i$ and a market portfolio with return $R_m = u_m + \beta_m Z + \sigma_m \sqrt{Z} N_m$. Let $\sigma_{im}$ be the correlation coefficient between the standard Normal variables $N_i$ and $N_m$. Suppose an investor allocates a fraction $\lambda$ of wealth to asset $i$ and the remaining fraction $(1-\lambda)$ to the market portfolio. The resulting portfolio return is $R_{\lambda} = \lambda R_i + (1-\lambda) R_m$. Its expected return is $
ER_{\lambda} = \lambda ER_i + (1-\lambda) ER_m,
$
and its (modified) standard deviation is
$
\bar{\sigma}(R_{\lambda}) = \sqrt{\lambda^2 c_i^2 + (1-\lambda)^2 \sigma_m^2 + 2\lambda(1-\lambda)\sigma_m c_i \sigma_{im}}.
$
We compute
$
\left.\frac{d ER_{\lambda}}{d\lambda}\right|_{\lambda=0} = ER_i - ER_m
$
and
$
\left.\frac{d \bar{\sigma}}{d\lambda}\right|_{\lambda=0}
= \frac{\sigma_m c_i \sigma_{im} - \sigma_m^2}{\sigma_m}.
$
Using the relation
\[
\left[\left.\frac{d ER_{\lambda}}{d\lambda}\right|_{\lambda=0}\right] \Big/
\left[\left.\frac{d \bar{\sigma}}{d\lambda}\right|_{\lambda=0}\right]
= \frac{ER_m - r_f}{\sigma_m},
\]
we obtain
\[
ER_i - r_f = \frac{c_i \sigma_{im}}{\sigma_m} (ER_m - r_f).
\]

By the definition of  $\rho$ we know that
\[
c_i = \frac{\rho(R_i) - ER_i}{\rho_0}
\quad \text{and} \quad
\sigma_m = \frac{\rho(R_m) - ER_m}{\rho_0}.
\]
Substituting these expressions yields
\begin{equation*}\label{skew1-capm}
ER_i - r_f = \sigma_{im}\,\frac{\rho(R_i) - ER_i}{\rho(R_m) - ER_m}(ER_m - r_f).
\end{equation*}
The equation (\ref{sigrm}) follows from $Cov(R, R_m)=b\beta_mVar(Z)+c\sigma_mEZ \sigma_{rm}$ and the equation for $\beta_{SI}$  follows from $Var(R_m)=\beta_m^2Var(Z)+\sigma_m^2EZ$.
\end{proof}

Observe that $\rho(R)-ER>0$ and $\rho(R_m)-ER_m>0$ always hold, since $\sqrt{Z}N$ is a symmetric random variable, which implies $\rho_0(\sqrt{Z}N)>0$. Furthermore, the tangent portfolio in Proposition \ref{4.33} can also be described as the optimizer of the Sharpe ratio
\[
\frac{ER_{\omega}-r_f}{\bar{\sigma}(R_{\omega})},
\]
where the risk measure specified in (\ref{new-risk}) is used.
\begin{remark} Recall that the standard CAPM is given by
\[
ER-r_f=\frac{Cov(R, R_m)}{Var{R_m}}(ER_m-r_f),
\]
where the covariance between the asset return $R$ and the market return $R_m$ appears explicitly. In contrast, the SI-CAPM (\ref{skew-capm}) does not contain this covariance term. Instead, the SI-CAPM is based on the correlation between $N_r$ and $N_m$. This correlation can be obtained from (\ref{sigrm}), or it can be calculated directly using $N_r=\frac{R-a+bZ}{c\sqrt{Z}}$ and $N_m=\frac{R_m-u_m-\beta_mZ}{\sigma_m\sqrt{Z}}$.

The traditional CAPM assumes normally distributed returns and uses variance as the key measure of risk, which leads to beta as the associated risk metric. Our relation (\ref{skew-capm}) delivers a similar linear connection between expected return and risk. Even when returns follow a heavy-tailed mixture distribution (for instance, a t-distribution) and risk is measured by a more sophisticated criterion (such as Expected Shortfall), the equilibrium pricing relation still takes the simple linear form
\[
ER=r_f+\beta (ER_m-r_f).
\]
In this setting, however, the beta $\beta$ is determined by the mixture model parameters rather than being driven solely by the covariance with the market.

As emphasized at the start of this Section, our SI-CAPM is developed under the assumption that all assets share the same mixing distribution almost surely. At first sight, this requirement that the mixing distribution be almost surely identical across assets may appear quite restrictive. However, when seeking a solid theoretical foundation for a widely applied model like the CAPM, the imposition of such strong assumptions seems practically unavoidable.
\end{remark}

\begin{remark}If $N_r$ and $N_m$ in Theorem \ref{capm-new} are perfectly correlated, then $\sigma_{rm}=1$, and in this situation the expression (\ref{skew-capm}) can be rewritten as
\[
\frac{ER-r_f}{\rho(R)-r_f}=\frac{ER_m-r_f}{\rho(R_m)-r_f},
\]
indicating that the ratio of the expected excess return $ER-r_f$ of any asset to its excess risk $\rho(R)-r_f$ coincides with the corresponding ratio for the market portfolio. If $\sigma_{rm}<1$, then we obtain
\[
\frac{ER-r_f}{\rho(R)-ER}=\sigma_{rm} \frac{ER_m-r_f}{\rho(R_m)-ER_m}.
\]
\end{remark}

\vspace{0.3in}

\textbf{Mean-$\rho$ frontier curve:} Next, we write down the corresponding  frontier curve. For this, we consider investments only on risky assets (\ref{one}). The return $R_{\omega}$ of a portfolio $\omega$ with $\omega^T\1=1$ is given by $R_{\omega}=\omega^T\mu+\omega^T\gamma Z+\sqrt{\omega^T\Sigma \omega}\sqrt{Z}N$. Since $\rho(R_{\omega})=\omega^T(\mu+\gamma EZ)+\sqrt{\omega^T\Sigma \omega}\rho_0$, the frontier portfolios are obtained by solving the following optimization problem 
\begin{equation}\label{frontier-app}
\begin{split}
\min_{\omega}\; &\omega^T\Sigma \omega,\\
\mbox{s.t.}\; &\omega^T(\mu+\gamma EZ)=r,\\
&\omega^T\1=1,\\
\end{split}    
\end{equation}
for all return levels $r$. We introduce the following notations:
\[
\mathcal{A}=(\mu+\gamma EZ)^T\Sigma^{-1} (\mu+\gamma EZ), \; \mathcal{B}=(\mu+\gamma EZ)^T\Sigma^{-1}\mathbf{1},\; \mathcal{C}=\mathbf{1}^T\Sigma^{-1}\mathbf{1}, \; \triangle=\mathcal{A}\mathcal{C}-\mathcal{B}^2.
\]
The solution of problem (\ref{frontier-app}) is given by 
\[
\omega_r=g+rh,
\]
where $g=\frac{1}{\triangle}[\mathcal{A}(\Sigma^{-1}\mathbf{1})-\mathcal{B}(\Sigma^{-1}\mu)]$ and $h=\frac{1}{\triangle}[\mathcal{C}(\Sigma^{-1}\mu)-\mathcal{B}(\Sigma^{-1}\mathbf{1})]$. One can easily calculate that
\[
\omega^T_r\Sigma \omega_r=\frac{1}{\triangle}(\C r^2-2\B r+\A).
\]
We know that
\[
\rho(R_{\omega_{r}})=\omega_{r}^T(\mu+\gamma EZ)+\sqrt{\omega_{r}^T\Sigma \omega_r}\,\rho_0,
\]
and, by construction, $\omega_{r}^T(\mu+\gamma EZ)=r$. Therefore, the frontier can be written in closed form as
\begin{equation}\label{m-r-frontier}
\rho=r+\frac{\rho_0}{\sqrt{\triangle}}\sqrt{\C r^2-2\B r+\A},
\end{equation}
in the $\rho$–$r$ coordinate plane. Note that since $\rho(\omega^TX)=\omega^T(\mu+\gamma EZ)+\bar{\sigma}(\omega^TX)\rho_0$ the frontier curve in (\ref{m-r-frontier}) is obtained from the frontier curve associated with the risk measure $\bar{\sigma}$ by an appropriate translation and rescaling. Furthermore, note that $\lim_{r\rightarrow -\infty}\rho(r)=+\infty$ provided that $\rho_0>\sqrt{\frac{\triangle}{\mathcal{C}}}$. Hence, in this situation, the frontier curve takes on a bullet-like shape.
 
We emphasize here that the mean-$\rho$ frontier  (\ref{m-r-frontier}) is a powerful and adaptable tool, generalizing the traditional mean–variance framework to work with any risk measure $\rho$ of the form (\ref{rhotilde}). In particular, it can reflect investor-specific notions of risk, deal with non-normal, heavy-tailed, and skewed return distributions, emphasize tail risk for regulatory or conservative investment objectives, and offer a solid theoretical basis for diversification benefits even when variance is not the chosen risk metric. By replacing a single, fixed risk measure with a generic $\rho$, the frontier evolves into a flexible framework suited to contemporary portfolio management, where risk is multi-dimensional and return distributions are typically non-normal.

\textbf{Skewness-induced frontier curve:} The frontier in the mean–$\bar{\sigma}$ plane is described by
\[
\bar{\sigma}=\sqrt{\C r^2-2\B r+\A}.
\]
According to Theorem \ref{efcvarrn}, this curve is obtained from mean–variance optimization using the return vector $\theta$ with mean $\mu_{\theta}=\mu+\gamma EZ$ and covariance matrix $\Sigma_{\theta}=\Sigma$. The curve is a hyperbola, and we obtain the straightforward relation
\[
\rho(r)=r+\frac{\rho_0}{\sqrt{\triangle}}\bar{\sigma}(r).
\]
In line with the findings of \cite{nur}, the utility-maximizing optimal portfolios are located on this frontier.

\textbf{Globally minimum risk portfolio:} The global minimum risk (GMR) portfolio is of substantial practical and theoretical relevance in finance. Its significance arises from its distinctive characteristics and its function in both portfolio design and financial economics. By construction, it is the risk-minimizing portfolio, and therefore the one whose future value an investor can forecast with the greatest degree of certainty. Whereas every other portfolio on the efficient frontier entails a balance between risk and expected return, the GMR portfolio is the one that most effectively reduces uncertainty. For an investor with strong risk aversion, this portfolio is therefore optimal. The global minimum risk portfolio is obtained by solving the following optimization problem:
\begin{equation}\label{gmr}
\begin{split}
\mbox{arg} \min_{\omega}\; &\rho(\omega^TX),\\
\mbox{s.t.}\; &\omega^T\1=1.\\
\end{split}    
\end{equation}
We denote the optimizer of (\ref{gmr}) by $\omega_{GMR}$. Note that we have $\rho(\omega^TX)=\omega^T(\mu+\gamma EZ)+\sqrt{\omega^T\Sigma \omega}\rho_0$. Consequently, problem (\ref{gmr}) reduces to an optimization problem involving functionals of quadratic forms, a type of problem analyzed in \cite{Zinoviy}. We decompose
\[
\Sigma=\begin{pmatrix}\Sigma_{11}& \Sigma_{12}\\  \Sigma_{21}& \sigma_{dd}\end{pmatrix},
\]
where $\Sigma_{11}$ is a $(d-1)\times (d-1)$ matrix and $\sigma_{dd}$ is the $(d,d)$-entry of $\Sigma$. Let $\1_1$ denote the $(d-1)$-dimensional column vector of ones. Define $m_1=(\mu_1+\gamma_1EZ, \cdots, \mu_{d-1}+\gamma_{d-1}EZ)^T$ and $ m_2=\mu_d+\gamma_dEZ$ (so that $\mu+\gamma EZ=(m_1, m_2)$). Set
\[
D=m_2\1_1-m_1, \; \; Q=\Sigma_{11}-\1_1\Sigma_{12}^T-\Sigma_{12}\1^T+\sigma_{dd}\1_1\1_1^T.
\]
Then, provided that $\rho_0>\sqrt{D^TQ^{-1}D}$ (see Theorem 1 in \cite{Zinoviy}), we obtain
\[
\omega_{GMR}=\frac{1}{\1^T\Sigma^{-1} \1}\Sigma^{-1}\1-\frac{1}{\sqrt{(\rho_0^2-D^TQ^{-1}D)(\1^T\Sigma \1)}}(D^TQ^{-1}, \1_1^TQ^{-1}D)^T.
\]

\textbf{Elliptical returns:} Consider now the special case $\gamma=0$ in model (\ref{one}). Then $X$ follows an elliptical distribution. In this setting, for any law-invariant coherent risk measure $\rho$, we obtain
\[
\rho(R_{\omega})=r_f+\omega^T(\mu-r_f\1)+\sqrt{\omega^T\Sigma \omega}\,\rho(\sqrt{Z}N).
\]
The market portfolio is then given by
\[
\omega_m=\frac{1}{\mathbf{1}^T\Sigma^{-1}(\mu-\mathbf{1} r_f)}\Sigma^{-1}(\mu-\mathbf{1} r_f).
\]
Let $\mathcal{D}_{\rho}$ be the deviation risk measure associated with $\rho$ (see \cite{Tyrrell} for details). Using arguments analogous to those above, we arrive at
\[
ER-r_f=\sigma_{rm}\frac{\mathcal{D}_{\rho}(R)}{\mathcal{D}_{\rho}(R_m)}(ER_m-r_f),
\]
or, equivalently,
\[
\frac{ER-r_f}{\mathcal{D}_{\rho}(R)}=\sigma_{rm}\frac{ER_m-r_f}{\mathcal{D}_{\rho}(R_m)},
\]
for any return of the form $R=a+c\sqrt{Z}N$. Under the additional assumption that $Z$ is square-integrable, we have $\sigma_{rm}=\frac{\text{Cov}(R, R_m)}{c\sigma_m EZ}$. Thus, in the case of elliptical (and in particular normal) distributions, the CAPM relation can be formulated in terms of any law-invariant coherent risk measure.

\vspace{0.1in}

\noindent \textbf{Acknowledgments}
\vspace{0.2in}

The author wishes to express gratitude to Alexander Schied for his comments and guidance during the initial phases of this project.

\bibliographystyle{plainnat}

\bibliography{main}

\end{document}